\documentclass[12pt,preprint]{aastex}
\usepackage[dvipdfm]{hyperref}
\usepackage{rotating}
\usepackage{lineno}
\linenumbers

\begin{document}
\title{Multi-wavelength Observations of Blazar \objectname{AO~0235+164} in the 2008-2009 Flaring State}
\shorttitle{Multi-wavelength Observations of \objectname{AO~0235+164}}
\author{
M.~Ackermann\altaffilmark{1},
M.~Ajello\altaffilmark{2},
J.~Ballet\altaffilmark{3},
G.~Barbiellini\altaffilmark{4,5},
D.~Bastieri\altaffilmark{6,7},
R.~Bellazzini\altaffilmark{8},
R.~D.~Blandford\altaffilmark{2},
E.~D.~Bloom\altaffilmark{2},
E.~Bonamente\altaffilmark{9,10},
A.~W.~Borgland\altaffilmark{2}
% ...
}
\altaffiltext{1}{Deutsches Elektronen Synchrotron DESY, D-15738 Zeuthen, Germany}
\altaffiltext{2}{W. W. Hansen Experimental Physics Laboratory, Kavli Institute for Particle Astrophysics and Cosmology, Department of Physics and SLAC National Accelerator Laboratory, Stanford University, Stanford, CA 94305, USA}
\altaffiltext{3}{Laboratoire AIM, CEA-IRFU/CNRS/Universit\'e Paris Diderot, Service d'Astrophysique, CEA Saclay, 91191 Gif sur Yvette, France}
\altaffiltext{4}{Istituto Nazionale di Fisica Nucleare, Sezione di Trieste, I-34127 Trieste, Italy}
\altaffiltext{5}{Dipartimento di Fisica, Universit\`a di Trieste, I-34127 Trieste, Italy}
\altaffiltext{6}{Istituto Nazionale di Fisica Nucleare, Sezione di Padova, I-35131 Padova, Italy}
\altaffiltext{7}{Dipartimento di Fisica ``G. Galilei", Universit\`a di Padova, I-35131 Padova, Italy}
\altaffiltext{8}{Istituto Nazionale di Fisica Nucleare, Sezione di Pisa, I-56127 Pisa, Italy}
\altaffiltext{9}{Istituto Nazionale di Fisica Nucleare, Sezione di Perugia, I-06123 Perugia, Italy}
\altaffiltext{10}{Dipartimento di Fisica, Universit\`a degli Studi di Perugia, I-06123 Perugia, Italy}
% ...

\clearpage
\begin{abstract}
The blazar \objectname{AO~0235+164} (z=0.94) has been one of the most
active objects observed by \textit{Fermi} Large Area Telescope (LAT) since its launch in Summer 2008.
In addition to the continuous coverage by \textit{Fermi},
contemporaneous observations were carried out from the radio to
$\gamma$-ray bands between 2008 September and 2009 February.
In this paper we summarize the rich multi-wavelength data
collected during the campaign (including F-GAMMA, GASP-WEBT, Kanata, OVRO, RXTE, SMARTS, Swift and other 
instruments), examine the cross-correlation
between the light curves measured in the different energy bands,
and interpret the resulting spectral energy distributions in the
context of well-known blazar emission models.  We find that the $\gamma$-ray
activity is well correlated with a series of near-IR/optical flares,
accompanied by the increase in the optical polarization degree.
On the other hand, the X-ray light curve shows a distinct 20-day
high state of unusually soft spectrum, which does not match the
extrapolation of the optical/UV synchrotron spectrum.
We tentatively interpret this feature as the bulk Compton
emission by cold electrons contained in the jet,
which requires an accretion disk corona with effective covering factor
of 19\% at a distance of 100 $R_{\rm g}$.  We model the broad-band
spectra with a leptonic model with external radiation dominated
by the infrared emission from the dusty torus.
\end{abstract}


\section{Introduction}

Blazars are a class of active galactic nuclei
characterized by high flux variability at all wavelengths and
compact (milli-arcsecond scale) radio emission of extreme brightness
temperatures, often exceeding the Compton limit \citep{key:urry}.  Their radio spectra
are generally well-described by a power-law shape, with a ``flat'' spectral
index $\alpha < 0.5$ (where the flux density $F_{\nu} \propto \nu^{-\alpha}$).
Multi-epoch VLBI (Very Long Baseline Interferometry) observations often show superluminal expansion, and
the radio and optical emission is usually highly polarized.
These general properties are well-described as
arising in a relativistic jet pointing close to our line of sight
\citep{key:BR78}.
The jet, presumably deriving its power
from accretion onto a supermassive, rotating black hole surrounded by an accretion
disk, contains ultrarelativistic electrons (with particle Lorentz factors
$\gamma_{\rm el}$ reaching $10^{3} - 10^{5}$, depending on the object).
These relativistic electrons produce soft photons from radio up to UV
(or in some cases, soft X-rays) through synchrotron emission, and high-energy
photons up to TeV energies, via the inverse-Compton process which involves
scattering of synchrotron photons (the SSC scenario), as well as scattering
of externally produced soft photons (the External Radiation Compton, ERC, scenario). A contribution to the
high energy radiation can also be provided
by synchrotron radiation of pair cascades powered by hadronic processes and
by synchrotron emission of ultra-high-energy protons and muons (see reviews
of radiative models of blazars by \citealt{key:SM01,key:levinson,key:boettcher}).
Noting difficulties of hadronic models to explain
the spectra of luminous blazars \citep{key:sikora09,key:sikora11},
we investigate in this paper  only leptonic models, i.e.,
the models which involve production of radiation by directly accelerated electrons.
Densely sampled, simultaneous monitoring observations
throughout the entire
electromagnetic spectrum from the radio to $\gamma$-ray bands can
provide important constraints on such models.

When emission lines are absent or weak, with an equivalent width (EW) less than {5~\AA}
in the rest frame \citep[see, e.g.,][]{key:stickel},
a blazar is classified as a BL Lac object;
otherwise it belongs to the class of flat-spectrum radio quasars (FSRQs).
While in a majority of BL Lac
objects - especially in those with the $\nu F_{\nu}$ spectral energy distribution
(SED) peaking in the far UV - to X-ray range (the so-called HSP, or ``high-synchrotron peaked
 BL Lac objects'') - detection of emission lines is
rare, and if detected, the lines are extremely weak (for recent measurements,
see, e.g., \citealt{key:stocke}), in the objects where the SED peaks in the infrared
or optical range (the so-called LSP, or ``low-synchrotron peaked BL Lac objects''),
easily discernible emission lines have been
detected often.  When detected, such lines provide a measurement of
redshift, but also yield crucial information about the details of
accretion in the central source.  In some cases such as \objectname{AO~0235+164} \citep{key:raiteri07},
discussed in this paper, and even BL Lacertae \citep{key:vermeulen,key:corbett}, the prototype of the BL Lac class,
the EW of emission lines can vary from one observational epoch to another.
This is
primarily due to large-amplitude variability of the nonthermal continuum, which becomes brighter
or fainter with respect to the presumably less-variable
emission lines.  Regardless, the
detailed properties of the emission lines are crucial in establishing
the radiative environment encountered by the jet emerging from the nucleus,
and thus are indispensable in establishing the most likely source of seed-photon
population for inverse Compton scattering.  While the most compelling scenario
has the
internal jet photons dominating this population  in the HSP sub-class, and
the external photons (from emission-line region, or disk photons rescattered
by the medium confining the lines) in FSRQs, the situation with
LSP BL Lac objects is unclear.

Studies of an LSP blazar \objectname{AO~0235+164} provide an exceptional opportunity to
answer this question.  It is one of the original BL Lac objects in the \cite{key:stein}
compilation, discovered via optical identification of a variable
radio source by \citet{key:spinrad}.   Early observations - as well as the
inspection of historical plates - revealed that optical variability can range over
5 magnitudes \citep{key:rieke}, motivating monitoring observations over a wide range
of frequencies since its discovery.
The redshift $z_{\rm em} = 0.94$ has been inferred from weak optical emission lines
by \cite{key:cohen87}, but even earlier optical spectroscopy revealed two absorption
line systems, one at $z_{\rm ab1} = 0.524$, and another, weaker one at $z_{\rm ab2} = 0.852$
discovered by \cite{key:burbidge} and by \cite{key:rieke}.  The intervening
$z_{\rm ab1} = 0.524$ system has also  been detected in absorption in the radio,
via the redshifted hydrogen 21 cm line by \cite{key:wolfe} and \cite{key:roberts},
but also as a Ly$\alpha$ absorber, revealing damped Ly$\alpha$ properties \citep{key:snijders},
and implying {\sl a considerable absorption in other bands}.  Detailed studies of that
absorbing system by \cite{key:junkkarinen} allow accurate
corrections to be applied to the observed optical spectra in order to determine reliably the intrinsic spectrum
of the blazar.
%and we use those corrections below.
Likewise, since the environment
in the field of \objectname{AO~0235+164} is complex and includes several possibly
interacting foreground galaxies at $z_{\rm ab1} = 0.524$ as well as the system
at $z_{\rm ab2} = 0.852$, the {\sl emission} in the optical-UV band (and to much
lesser degree, in the soft X-ray band) may be contaminated.  One galaxy, probably
a normal spiral, is 1.3 arcsec east, while another object, about 2~arcsec to the
south, is known to be an AGN and could affect the flux of \objectname{AO~0235+164}
when it is very faint, especially in the bluer part of the spectrum \citep{key:raiteri2005}.
%We account for those as well in our analysis below.

Historical data for this source are abundant.  Radio observations were performed
by many instruments, starting from about 100~MHz up to 300~GHz, and including
multi-epoch VLBI studies \citep{key:jorstad}. Space and ground-based infrared data
are available from sub-mm (far-IR) down to micron wavelengths (near-IR);
optical bands,  UBVRI,  have been extensively monitored by many telescopes around
the world. \objectname{AO~0235+164} has also been detected in the high energy band by
essentially all soft X-ray observatories including $Einstein$ (\citealt{key:einstein}),
$EXOSAT$ (\citealt{key:exosat}), $ROSAT$ (\citealt{key:madejski}, \citealt{key:rosat}), $ASCA$
(\citealt{key:madejski}, \citealt{key:junkkarinen}), {\it Beppo-SAX} (\citealt{key:bepposax}),
$RXTE$ (\citealt{key:rxte}), and {\it XMM-Newton} (\citealt{key:raiteri2008}).  This source has also been
identified as a powerful and strongly variable $\gamma$-ray emitter via observations
by $EGRET$ onboard the {\it Compton Gamma-Ray Observatory} ($CGRO$) in the high $\gamma$-ray
energy range from 30~MeV to 20~GeV, with six pointings between 1992 and 1997 providing
two detections \citep{key:hunter,key:madejski} and four upper limits.  The mid-energy
$\gamma$-ray emission was probed by {\it COMPTEL} during $CGRO$ Cycle 4 (1994-1995),
yielding only upper limits for the flux in the interval of
0.75-30~MeV.  These numerous multi-wavelength observations show that
\objectname{AO~0235+164} is characterized by extreme variability on long
(month-years) and short (intraday) time scales over a wide range of
the electromagnetic spectrum.

The study of blazars, of their broad-band spectra and of their complex variability,
has been greatly enriched since the start of scientific observations with the \textit{Fermi}
Large Area Telescope (LAT)
in 2008 August \citep{key:Atwood_LAT} thanks to its high
sensitivity and essentially uninterrupted observations afforded by the survey mode.
Such new and sensitive $\gamma$-ray observations motivated many multi-band campaigns,
often conducted with dedicated facilities, and \objectname{AO~0235+164} was (and
continues to be) one of the well-sampled targets.
This paper presents the results
of the LAT monitoring of \objectname{AO~0235+164}, as reported in Section
\ref{sec:Fermi}. The description of multi-wavelength observations
conducted between 2008 August and 2009 February when the source showed strong
activity in $\gamma$-rays as well as in radio through optical and X-ray
bands \citep{key:atel1744,key:atel1784}, follows in Section \ref{sec:mw_data}.
The analysis of those data, including the discussion of the temporal profiles measured
in various bands and the connection to the $\gamma$-ray activity,
is reported in Section \ref{sec:mw_lightcurve}.
A significant part of these data have been independently analyzed by \cite{key:agudo11}.
In Section \ref{sec:sed} we present the overall spectral
energy distribution (SED) and its temporal behavior, and discuss the implications
of the data on the modeling of emission processes and the structure of the jet in
\objectname{AO~0235+164}:  there, we argue that while the equivalent width
of emission lines in this object might suggest a classification as a
BL Lac object, the isotropic luminosity inferred from the data indicates
it is a quasar.
In Section \ref{sec:modeling}, we show models of the broad-band
emission in the context of synchrotron + Compton models.  Our
consideration of the broad-band SED
suggests that the most likely mechanism for $\gamma$-ray emission is
Comptonization of circumnuclear IR radiation from dust, commonly
present in quasars.
This is a different scenario from the one proposed by \cite{key:agudo11}, who argued for the synchrotron self-Compton process. We discuss these two approaches in Section \ref{sec:discussion}.
We conclude with a summary of our results in Section \ref{sec:conclusions}.

\section{{\textit Fermi}-LAT Observations and Data Analysis}
\label{sec:Fermi}

The LAT, the primary instrument onboard the \textit{Fermi} $\gamma$-ray observatory,
is an electron-positron
pair conversion telescope sensitive to $\gamma$-rays of energies from
20~MeV to $>$ 300~GeV. The LAT consists of a high-resolution silicon
microstrip tracker, a CsI hodoscopic electromagnetic calorimeter and
an anticoincidence detector for the identification of charged particles background.
The full description of the instrument and its performance can be found
in \cite{key:Atwood_LAT}. The large field of view ($\sim$2.4~sr) allows
the LAT to observe the full sky in survey mode every 3~hours. The LAT point
spread function (PSF) strongly depends on both the energy and the
conversion point in the tracker, but less so on the incidence angle.
For 1~GeV normal-incidence conversions in the upper section of the
tracker the PSF 68$\%$ containment radius is 0.8$^{\circ}$.

The \textit{Fermi}-LAT data of \objectname{AO~0235+164} presented here were
obtained in the time period between 2008 August and 2009 February
when \objectname{AO~0235+164}
entered a bright high $\gamma$-ray state;
and immediately after, dropped to lower states. The data have been
analyzed by using the standard \textit{Fermi}-LAT software
package\footnote{\url{http://fermi.gsfc.nasa.gov/ssc/data/analysis/documentation/Cicerone/}}.
The Pass 6 Diffuse event class and {\tt P6$\_$V3$\_$DIFFUSE} instrument response functions
(Atwood et al. 2009) were used in our analysis. We selected events within a 15$^{\circ}$
region of interest (RoI) centered on the source position, having energy greater
than 100 MeV. The data have been analyzed using the Science
Tools software package (version v9r16).
In order to avoid background contamination from the bright Earth limb,
time intervals when the Earth entered the LAT Field of View were
excluded from the data set. In addition, events with zenith angles larger
than 105$^{\circ}$ with respect to the Earth reference frame
\citep{key:bright_source_list} were excluded from the analysis. The data were
analyzed with an unbinned maximum likelihood technique described by \cite{key:mattox}
using the analysis software ({\tt gtlike}) developed by the LAT
team\footnote{\url{http://fermi.gsfc.nasa.gov/ssc/data/analysis/documentation/Cicerone/Cicerone\_Likelihood}}.

Accurate spectral and flux measurements require a reliable accounting
for the diffuse foreground due to the Galactic interstellar emission, as well as
the extragalactic diffuse $\gamma$-ray emission, the residual cosmic ray
background, and contamination from nearby sources.  The fitting procedure
simultaneously fits for the parameters of the source of interest as well as
of nearby $\gamma$-ray sources and the diffuse backgrounds, which
in turn have been modeled using
{\tt gll$\_$iem$\_$v02} for the Galactic diffuse emission
and {\tt isotropic$\_$iem$\_$v02} for the extragalactic
isotropic emission models\footnote{\url{http://fermi.gsfc.nasa.gov/ssc/data/access/lat/BackgroundModels.html}}.

The sources surrounding \objectname{AO~0235+164} were modeled using a
power-law function:
\begin{equation}
\frac{dN}{dE} = \frac{N(1-\Gamma)E^{-\Gamma}}{E_{max}^{1-\Gamma}-E_{min}^{1-\Gamma}}
\end{equation}
where N is the normalization factor and $\Gamma$ the photon index.
In the fitting procedure, all sources
within 10$^{\circ}$ were included in the model with the normalization factor N free, while the sources
located between 10$^{\circ}$ and 20$^{\circ}$ had all the model parameters fixed
to the 1FGL catalog values \citep{key:1FGLcatalog}.

The plots in subsequent sections show only statistical errors for the fit parameters.
Systematic errors arise mainly from uncertainties on the LAT effective area, which is
derived from the on-orbit estimations. These errors could be as large as 10$\%$
below 0.1~GeV, $<$5$\%$ near 1~GeV and 20$\%$ above 10~GeV.

\subsection{$\gamma$-ray Light Curve}
\label{sec:gammalc}

The light curve of \objectname{AO~0235+164} in the \textit{Fermi}-LAT energy range has been
assembled using 3-day long time bins and covers the first 6 months of data taking
from 2008 August 4 to 2009 February 4, when the source was in a high state and
a large set of multi-wavelength observations is available.
The light curve is obtained applying the {\tt gtlike} fit
across the overall energy range considered, from 100~MeV to 100~GeV, in each of the selected time bins.
For each time interval the flux and the photon index
of \objectname{AO~0235+164} are determined using the maximum likelihood algorithm
implemented in {\tt gtlike}, following the procedure outlined in the previous section.
The data
are modeled with a power-law function with both the normalization factor %$N$
and photon index %$\gamma$
left free in the likelihood fit.

\begin{figure}[htbp]
\begin{center}
\includegraphics[width=1.0\textwidth]{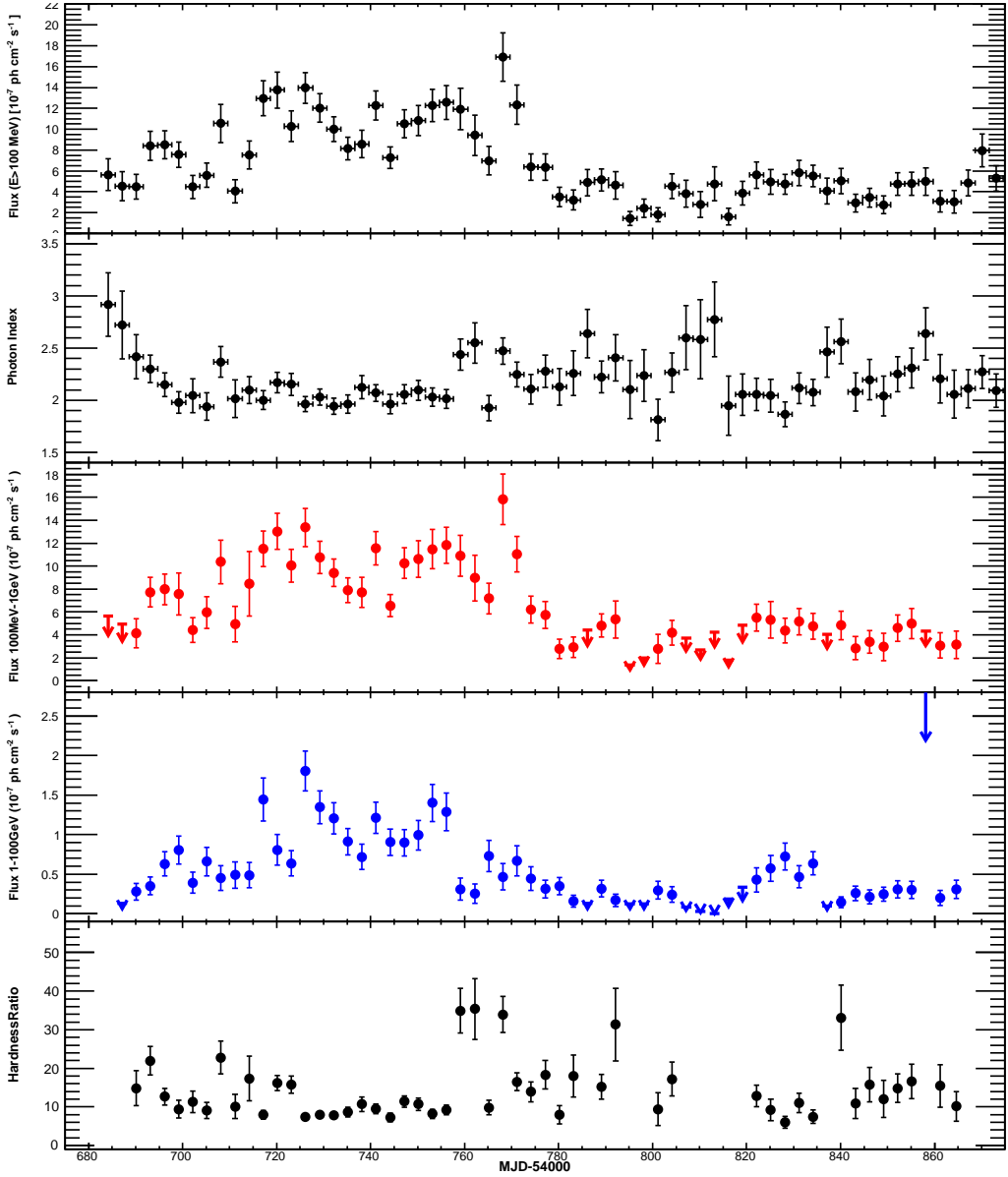}
\caption{\textit{Fermi}-LAT light curve from 2008 August 4 to 2009 February 4 in 3-day time intervals.
The first panel from the top shows the flux,  in the energy range from 100~MeV to 100~GeV,  derived from the 
{\tt gtlike} fit in the 3-days time intervals, 
assuming a simple power law spectrum.
The second panel shows the photon index $\Gamma$ in the same energy range from 100~MeV to 100~GeV.
The third panel shows the light curve evaluated in the energy
range from 100~MeV to 1~GeV.
The fourth panel shows the light curve in the energy
range from 1~GeV to 100~GeV.
The last panel shows the hardness ratio defined as $\frac{F_{100MeV-1GeV}}{F_{1GeV-100GeV}}$ for the data points having a TS$>$10 and F$_{err}$(E)/F(E)$>$0.5 in both energy ranges.
The hardness ratio is not evaluated if either of the two fluxes is an upper limit.}
\label{fig:Fermi_LC}
\end{center}
\end{figure}

The 6-month $\gamma$-ray light curve is reported in Figure~\ref{fig:Fermi_LC}
together with the photon index resulting from the likelihood fit in each time bin.
The trend in the entire energy range from 100~MeV to 100~GeV shows a clear
high-state period followed by a final, narrow, high-flux peak.

The temporal behavior of the source in $\gamma$-rays was also
studied in two separate energy ranges, from 100~MeV to 1~GeV and
from 1~GeV to 100~GeV and the hardness ratio among the two bands has been
determined.  The analysis follows the same procedure described above
to determine the overall light curve and the results are shown in
the three bottom panels of Figure~\ref{fig:Fermi_LC}.
%The top and middle panels show the \textit{Fermi}
%light curves evaluated in 100~MeV-1~GeV  and  1~GeV-100~GeV
%energy ranges respectively.

The arrows in the light curves represent 95\% upper limits, which are calculated for
data points with a test statistic (TS)\footnote{The Test Statistics
is defined as TS = -2$\times$(log(L$_1$)-log(L$_0$)) with L$_0$ the
likelihood of the Null-hypothesis model as compared to the likelihood of
a competitive model, L$_1$;  see Mattox et al. (1996).} lower than 10 (which corresponds to
a significance somewhat higher than 3$\sigma$), or with a value of the ratio between
flux error and flux (F$_{err}$(E)/F(E))$\geq$0.5 in order to obtain meaningful data points.

The results show that both the low- and high-energy profiles follow
the same trend. Nevertheless, it is interesting to underline that
the narrow peak at the end of the high-state period is mainly due
to an enhanced low-energy flux.  The ratio among the two fluxes also shows a
value higher than the average in the same time interval.

% \begin{figure}[htbp]
% \begin{center}
% \includegraphics[width=1.0\textwidth]{fig02.eps}
% \caption{Top panel: \textit{Fermi}-LAT light curve from 2008 August 4
% to 2009 February 4 in 3-day time intervals evaluated in the energy
% range from 100~MeV to 1~GeV. Middle panel: \textit{Fermi}-LAT light curve
% in the same period and for the same time bins evaluated in the energy
% range from 1~GeV to 100~GeV. Bottom panel: Hardness ratio defined as
% the ratio $\frac{F_{100MeV-1GeV}}{F_{1GeV-100GeV}}$ for the data
% points having a TS$>$10 and F$_{err}$(E)/F(E)$>$0.5 in both energy ranges. The hardness ratio is not evaluated if either of the two fluxes is an upper limit.}
% \label{fig:hr}
% \end{center}
% \end{figure}

\subsection{$\gamma$-ray Spectral Analysis}
\label{sec:gammaspectrum}
\begin{figure}[htbp]
\begin{center}
\includegraphics[width=1.0\textwidth]{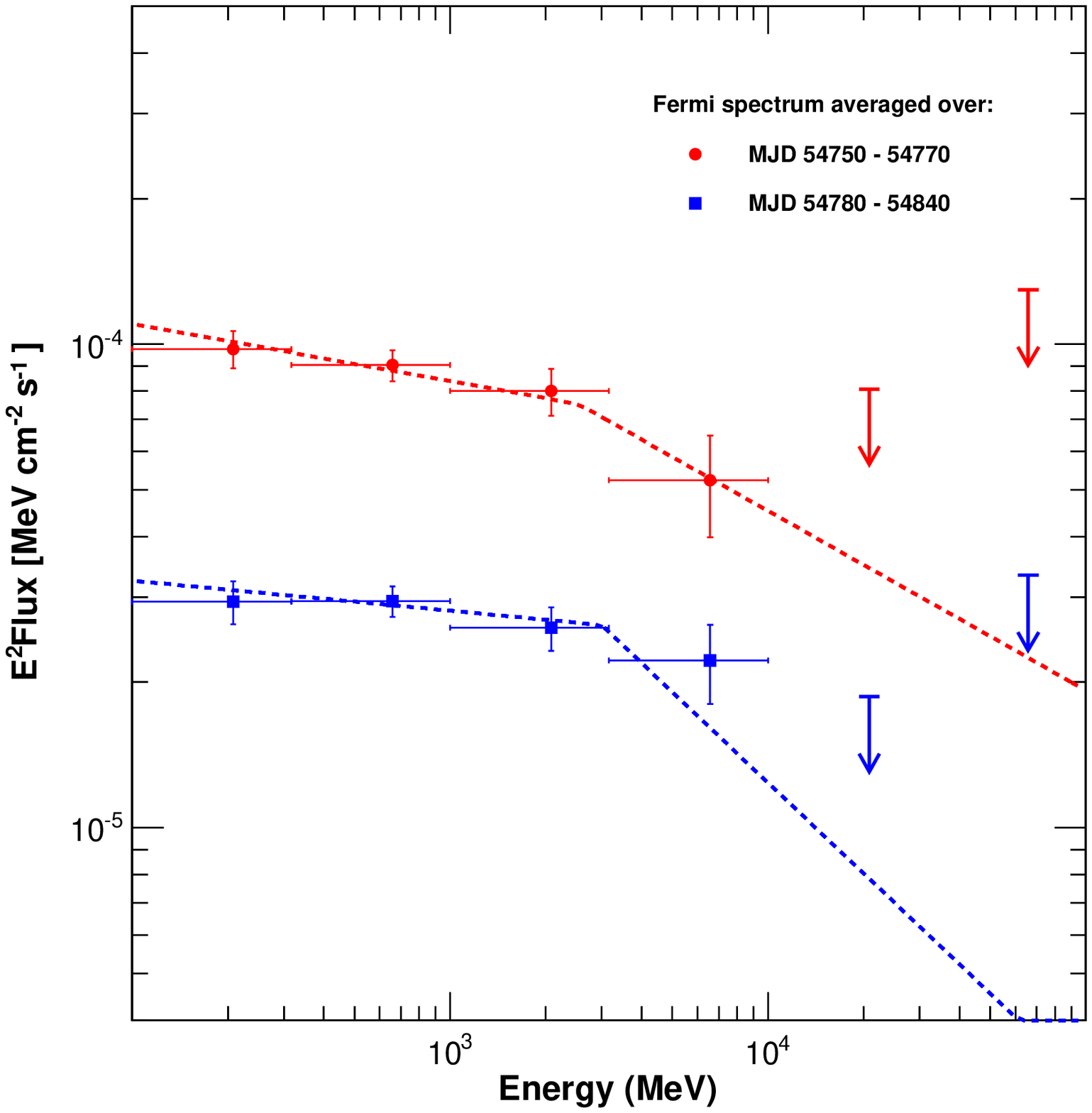}
\caption{\textit{Fermi}-LAT energy spectrum evaluated in different
time intervals corresponding to the X-ray flare (red circles)
from MJD 54750-54770 and $\gamma$-ray low state (blue squares) from MJD 54780-54840.}
\label{fig:Fermispectrum}
\end{center}
\end{figure}

The unbinned {\tt gtlike} analysis has been applied to
produce the $\gamma$-ray energy spectra shown
in Figure~\ref{fig:Fermispectrum}.  There, we divided the
full energy range from 100~MeV to 100~GeV
into 2 equal logarithmically spaced bins per decade.
In each energy bin a TS value greater than 10 and a
ratio between flux error and flux lower than 0.5 was required to quote a flux in that band,
otherwise a 95\% upper limit was given.

The standard {\tt gtlike} tool was applied in each energy bin,
modeling all the point sources in the region with a simple power-law spectrum with
photon index fixed to 2.
The normalization parameters of all point-like
sources within 10$^{\circ}$ were left as free parameters in the
fitting procedure, while the diffuse background components were
modeled as described above in \ref{sec:gammalc}.
Two time intervals were selected for the $\gamma$-ray spectral
analysis: the first corresponding to the X-ray flare interval
(MJD 54750-54770), the second associated with the
subsequent low $\gamma$-ray state (MJD~54780-54840).
In those time intervals, both power-law and broken-power-law functions provide a good fit of the spectral data.
We show the results of the broken-power-law fit, since
it provides a better fit to the high-energy spectrum of the source, from 100~MeV to
100~GeV, than a simple power law on the larger time intervals, as already studied in detail by \cite{key:spectralpro}.

\begin{equation}
 \frac{dN}{dE} = N_0\times\left\lbrace
\begin{array}{ccc}
(E/E_{b})^{-\Gamma_{1}} &; & if E<E_{b} \\
(E/E_{b})^{-\Gamma_{2}} &;& otherwise
\end{array}
\right.
\end{equation}

\begin{table}[th]
\begin{center}
\begin{tabular}{ccccc}
\hline
Time Interval & Flux & $\Gamma_1$ & $\Gamma_2$ & Break Energy \\
MJD & 10$^{-7}$ ph cm$^{-2}$s$^{-1}$ & & & GeV \\
\hline
\hline
54750-54770 & 11.05$\pm$1.48 & 2.12$\pm$0.16 & 2.37$\pm$0.32 & 2.6$\pm$0.8 \\
54780-54840 & 3.42$\pm$0.65 & 2.07$\pm$0.17 & 2.77$\pm$0.32 & 3.8$\pm$1.2 \\
\hline
\end{tabular}
\caption{Results of the {\tt gtlike} fit of the $\gamma$-ray spectrum during the high and low states.}
\label{tab:fitresults}
\end{center}
\end{table}

In both time intervals the $\Gamma_1$ index remains stable, while $\Gamma_2$ increases
showing a softening of the high energy part of the spectrum when the source is in
a fainter state, when also an increase in the break energy is observed. As also can be seen
in Figure~\ref{fig:Fermi_LC}, the high $\gamma$-ray state around MJD 54760 is essentially due
to the low energy photons ($<$1~GeV) and the spectra in Figure~\ref{fig:Fermispectrum}
show that the relative difference between the E$^2$Flux values above 1~GeV and below
1~GeV is higher in the time interval around the flare than during the low $\gamma$-ray state.

\section{Multi-wavelength Observations and Data Analysis}
\label{sec:mw_data}

The multi-wavelength campaign conducted on \objectname{AO~0235+164} in 2008-2009
saw a wide international participation. Table~\ref{tab:mwsummary} reports the
list of participating observatories, the energy bands, the period of
observation and the number of collected data points.

\begin{table}[hp]
\tiny
\begin{center}
\begin{tabular}{lllll}
\hline
\multicolumn{2}{l}{Observatory}  & Bands & Period of Observation & Data Points \\
\hline
\hline
\multicolumn{5}{c}{\textit{\textbf{Radio}}} \\
\textbf{GASP-WEBT} & Mauna Kea (SMA), USA & 345 GHz & 54664-54840 & 10 \\
& & 230 GHz & 54645-54842 & 23 \\
& Medicina, Italy & 5~GHz & 54724 & 1 \\
& & 8~GHz & 54606-54777 & 7 \\
& & 22~GHz & 54604-54779 & 7 \\
& Mets\"{a}hovi (KURP-GIX),Finland  & 37 GHz & 54633-54839 & 54 \\
& Noto, Italy & 43~GHz & 54642-54841 & 8 \\
& UMRAO, USA & 5 GHz & 54677-54840 & 16 \\
& & 8 GHz & 54621-54851 & 20 \\
& & 14.5 GHz & 54633-54848 & 41 \\
\textbf{F-GAMMA} & Effelsberg 100-m & 2.64 & 54414-55227 & 23 \\
& & 4.85~GHz & 54414-55227 & 24 \\
& & 8.35~GHz &  54414-55227 & 24 \\
& & 10.45~GHz & 54414-55227 & 25  \\
& & 14.6~GHz &  54422-55227 & 22 \\
& & 23.05~GHz & 54422-55227 & 17 \\
& & 32~GHz & 54616-55227 & 11 \\
& & 42~GHz & 54546-55227 & 8 \\
& IRAM 30-m & 86.2~GHz & 54382-55228 & 17 \\
& & 142.3~GHz & 54382-55228 & 15 \\
& & 228.4~GHz & 54440-54806 & 5 \\
\multicolumn{2}{l}{\textbf{IRAM PdBI}} & 88.9~GHz & 54700 & 1 \\
& &  169~GHz & 54883 & 1 \\
\multicolumn{2}{l}{\textbf{OVRO}} & 15 GHz & 54661-54848 & 49 \\
\hline
\multicolumn{5}{c}{\textit{\textbf{Near-Infrared}}} \\
\textbf{GASP-WEBT}  & Campo Imperatore & J & 54645-54794 & 87\\
&  & H & 54645-54794 &  82 \\
&  & K &  54645-54794 &  83 \\
\multicolumn{2}{l}{\textbf{Kanata}} & J &  54690-54753 & 21 \\
& & V &  54690-54753 & 21 \\
& & Ks &  54690-54753 & 21 \\
\multicolumn{2}{l}{\textbf{SMARTS}} & J & 54662-54847 & 69 \\
& & K &  54662-54842 & 39 \\
\hline
\multicolumn{5}{c}{Optical} \\
\textbf{GASP-WEBT} & Abastumani 70~cm & R & 54687-54780 & 287 \\
& Armenzano, 40~cm & R & 54699-54727 & 16 \\
& Calar Alto & R & 54712-54887 & 5 \\
& Crimean 70cm; ST-7 & R & 54691-54805 & 218 \\
& Kitt Peak (MDM 130~cm) & R & 54745-54801 & 50 \\
& L'Ampolla & R & 54778-54784 & 2 \\
& Lulin (SLT) & R & 54688-54862 & 120 \\
& Roque (KVA 35~cm) & R & 54748-54862 & 29 \\
& San Pedro Martir 84~cm & R & 54709-54773 & 15 \\
& St. Petersburg & R & 54698-54865 & 41 \\
& Talmassons & R & 54728-54843 & 11 \\
& Tuorla & R & 54722-54732 & 3 \\
\multicolumn{2}{l}{\textbf{SMARTS}} & R & 54662-54868 & 71 \\
& & B & 54662-54871 & 69 \\
& & V & 54662-54859 & 68 \\
\multicolumn{2}{l}{\textbf{Steward}} & R & 54743-54832 & 39 \\
& & V & 54743-54863 & 44 \\
\multicolumn{2}{l}{\textbf{SWIFT-UVOT}} & U & 54711-54818 & 16 \\
& & B & 54711-54818 & 16 \\
& & V & 54711-54818 & 16 \\
\hline
\multicolumn{5}{c}{Ultra-Violet} \\
\multicolumn{2}{l}{\textbf{SWIFT-UVOT}} & UVW1 & 54711-54818 & 16 \\
& & UVM2 & 54711-54818 & 15 \\
& & UVW2 & 54711-54818 & 16 \\
\hline \hline
\end{tabular}
\caption{Observatories participating in the work, periods of observations and number of data points used in this analysis.}
\label{tab:mwsummary}
\end{center}
\end{table}

\subsection{Effect of intervening material in the line of sight
on the optical, UV and X-ray data}
\label{sec:extinction}

Conversion of the observed optical magnitudes into the intrinsic flux
densities requires a special care, because the source emission is absorbed
not only in our Galaxy, but also by the elliptical galaxy in the line
of sight at redshift $z=0.524$, as outlined in the Introduction.
\cite{key:junkkarinen} tried several different extinction models for AO~0235+164, concluding that the best fit to their HST/STIS data is obtained by using models of \cite{key:cardelli} with $R_{\rm V}=3.1$ and $E_{\rm B-V}=0.154$ for the Galaxy, and $R_{\rm V} = 2.51$ and $E_{\rm B-V} = 0.227$ for the $z=0.524$ system.
This model accurately reproduces the $2175\AA$ absorption feature produced by the $z=0.524$ galaxy, but the far-UV end of their spectrum indicated a sharp hardening.
\cite{key:raiteri2005} proposed that this far-UV hardening is real and that it marks the onset of a new spectral component.
However, the fact that the shape of this feature does not change with the overall optical/UV luminosity indicates that the whole optical/UV spectrum is produced by a single synchrotron component, which intrinsic shape must be close to a power-law.
The far-UV hardening most likely is an artefact of overestimated extinction from the dust in the $z=0.524$ galaxy.
We modify the best-fit extinction model of \cite{key:junkkarinen} by replacing the \cite{key:cardelli} model for the $z=0.524$ galaxy with an analytical model of \cite{key:pei}.
In the first step, we modify the ``Milky Way'' model with parameters listed in Table 4 of \cite{key:pei} to match the \cite{key:cardelli} model for $R_{\rm V} = 2.51$ and $E_{\rm B-V} = 0.227$.
In particular, we adopt $\lambda_{\rm 2175A}=2170\AA$ and $n_{\rm FUV}=5.5$, and we multiply the normalization parameters $a_i$ by additional factors $f_i$: $f_{\rm FUV}=1.5$, $f_{\rm 2175A}=1.33$ and $f_{\rm BKG}=1.05$.
In the second step, we turn off the ``FUV'' component of the \cite{key:pei} model for the $z=0.524$ galaxy by setting $f_{\rm FUV}=0$.
This modification affects only the observed wavelengths shorter than $\sim 3300\AA$, the location of the $2175\AA$ feature redshifted by $z=0.524$, and is necessary to align the FUV spectra with the optical-NIR spectra (see Figure \ref{fig:sed_overall}).
We stress that extinction at longer wavelengths is very well constrained by the clear detection of a redshifted $2175\AA$ feature by \cite{key:junkkarinen}, and thus cannot be increased.
The resulting total extinction values $A_{\lambda}$ for the Swift/UVOT filters are: W2: 2.87; M2: 2.94; W1: 2.52; U: 2.71; B: 1.84; V: 1.46.
For the remaining optical and near IR filters, we use the values from Table 5 of \cite{key:raiteri2005}: R: 1.26; I: 0.90; J: 0.46; H: 0.28; K: 0.17.
We calculate the incident flux
$F_{\rm inc, \lambda}$ in the band corresponding to $\lambda$ from
the observed (absorbed) flux $F_{\rm abs, \lambda}$ via
$F_{\rm abs, \lambda} / F_{\rm inc, \lambda} = 10^{A_{\lambda}/2.5}$.
The same corrections are applied to the ground-based optical data.

% is ELISA the same galaxy as the z=0.524 absorbing system?
In addition, the source photometry is contaminated by the emission
of a nearby AGN (named ELISA by \cite{key:raiteri2005}). Hence, we
subtracted the ELISA contribution from the observed flux densities and
then corrected for the combined extinction of both galaxies, following
the prescriptions given by \cite{key:raiteri2005} and \cite{key:raiteri2008}.

Likewise, the X-ray data need to be corrected for the effect of
absorption:  here, the absorption effects
of {\sl both} our own Galaxy and the
intervening $z = 0.524$ system are considerable.
\cite{key:madejski} and \cite{key:junkkarinen} argue that the
absorption in the intervening system  originates in material with abundances
different from Galactic and, in reality, correct modeling of such
absorption should take this into effect.  However, as discussed by \cite{key:madejski}, the combined $ROSAT$ and $ASCA$ spectral fitting suggests that
this effect is relatively modest, the joint $ROSAT-PSPC$ and $ASCA$ data are adequately fitted by an absorbing column of
$2.8 \pm 0.4 \times 10^{21}$ cm$^{-2}$ located at $z=0$.  Since
the \textit{Swift} XRT data
have somewhat lower signal-to-noise ratio $(S/N)$ than the $ASCA$ observations, we simply adopt such a
``local'' model for absorption, since the main objective of our observations
was to determine the underlying continuum  of the \objectname{AO~0235+164}
rather than the detailed spectral properties of the absorber.  We note that this
value is in fact consistent with the spectral fit to the \textit{Swift} XRT data.

\subsection{GASP-WEBT}
The GLAST-AGILE Support Program (GASP) of the Whole Earth Blazar
Telescope (WEBT) was initiated in 2007 with the aim of performing a
long-term multi-wavelength monitoring of bright, $\gamma$-loud blazars
\citep{key:villata2008a,key:villata2009, key:dammando2009,key:raiteri2010}.
The GASP optical ($R$ band), near-IR,
and radio data are intended to complement the high-energy
observations by the $AGILE$ and \textit{Fermi} (formerly $GLAST$) satellites.

\objectname{AO~0235+164} has been the target of several WEBT
campaigns in the past \citep{key:raiteri2001,key:raiteri2005,key:raiteri2006,key:raiteri2008} and it is now one of the GASP sources
of highest observing priority.  During the high $\gamma$-ray state observed in the second half of 2008, the
source underwent an exceptional optical-to-radio outburst
closely monitored by the GASP \citep{key:villata2008b,key:villata2008c,key:bach2008}.  The GASP optical
data presented here were taken at the following observatories:
Abastumani, Armenzano, Calar Alto,
Crimean, Kitt Peak (MDM), L'Ampolla, Lulin, Roque de los Muchachos
(KVA), San Pedro Martir, St. Petersburg, Talmassons, and Tuorla.
Near-IR data in the $J$, $H$, and $K$ bands are all from Campo Imperatore.
Millimeter and centimeter radio observations were performed at the
SMA (230 and 345 GHz), Noto (43 GHz), Mets\"ahovi (37 GHz),
Medicina (5, 8, and 22 GHz), and UMRAO (4.8, 8.0, and 14.5 GHz) observatories.
All IR, optical, and UV data are corrected for the effects of the
intervening absorber (both due to the Milky Way, and the intervening galaxy)
as outlined above.

\subsection{F-GAMMA}
During the 2008-2009 flaring period, quasi-simultaneous multi-frequency cm/mm-band (from 2.64\,GHz to 230\,GHz) observations of \objectname{AO~0235+164}
were obtained using the Effelsberg 100-m and IRAM 30-m telescopes, within the framework of a \textit{Fermi} related monitoring
program of $\gamma$-ray blazars (F-GAMMA
program\footnote{\url{http://www.mpifr-bonn.mpg.de/div/vlbi/fgamma/fgamma.html}},
\cite{key:fuhrmann2007}, \cite{key:angelakis2008}).

The Effelsberg measurements were conducted with the secondary focus heterodyne
receivers at 2.64, 4.85, 8.35, 10.45, 14.60, 23.05, 32.00 and 43.00\,GHz.
The observations were performed quasi-simultaneously with cross-scans,
by slewing over the source position in the azimuth and elevation
directions with an adaptive number of sub-scans chosen to reach the desired
sensitivity (for details, see \citealt{key:fuhrmann2008}; \citealt{key:angelakis2008}).
Consequently, pointing offset correction, gain correction,
atmospheric opacity correction and sensitivity correction have been
applied to the data.

The IRAM 30-m observations were carried out with calibrated cross-scans
using the single pixel heterodyne receivers B100, C150, B230 operating
at 86.2, 142.3 and 228.4\,GHz. The opacity corrected intensities were converted
into the standard temperature scale and finally corrected for small remaining
pointing offsets and systematic gain-elevation effects. The conversion to
the standard flux density scale was done using the instantaneous conversion
factors derived from frequently observed primary (Mars, Uranus) and secondary
(W3(OH), K3-50A, NGC\,7027) calibrators.

\subsection{OVRO}
Observations of \objectname{AO~0235+164} at 15~GHz with the Owens Valley Radio
Observatory (OVRO) 40-meter telescope were made as part of an ongoing
blazar monitoring program \citep{key:richards2011}.  The 40-m telescope
is equipped with a cooled receiver at the prime focus, with a 3.0~GHz
bandwidth centered on 15.0~GHz and 2.5~GHz noise-equivalent reception
bandwidth.  The receiver noise temperature is about 30~K, and the
total system noise temperature including CMB, atmospheric, and ground
contributions is about 55~K.  A dual off-axis corrugated horn feed
projects two approximately Gaussian beams (157 arcsec full width half maximum, FWHM) on the
sky, separated in azimuth by 12.95 arcmin.  Dicke switching between
the two beams is performed using the cold sky in the off-source beam
as a reference, and a second level of switching is performed by
alternating the source between the two beams to cancel atmospheric and
ground noise.  Calibration is achieved using a stable diode noise
source for relative calibration and is referred to observations of
3C~286, for which we assume a flux density of 3.44~Jy \citep{key:Baars1977}
 with about 5\% absolute scale error.  OVRO flux density
measurements have a minimum uncertainty of 4~mJy in 32~s of on-source
integration, and a typical RMS relative error of 3\%.

\subsection{IRAM Plateau de Bure Interferometer (PdBI)}

The Plateau de Bure Interferometer (PdBI; \cite{key:winters2010}) is
able to observe in three atmospheric windows located around wavelengths
of 1.3~mm, 2~mm, and 3~mm. Each of these bands covers a continuous range
of frequencies that are available for observations; these ranges are
201--267~GHz for the 1.3~mm band, 129--174~GHz for the 2~mm band,
and 80--116~GHz for the 3~mm band.

Systematic monitoring of AGN is a by-product of regular observatory operations.
The PdBI uses active galactic nuclei as phase and amplitude calibrators.
Usually, one or two calibrators are measured every $\sim$20 minutes for $\sim$2 min (per source) throughout an observation. Antenna
temperatures are converted into physical flux densities using
empirical antenna efficiencies as conversion factors. These factors
are functions of frequencies and are located in the range from
$\sim$22~Jy/K (for the 3-mm band) to $\sim$37~Jy/K (for the 1.3-mm band).

The PdBI is equipped with dual linear polarization Cassegrain focus receivers.
This makes it possible to observe both orthogonal polarizations -- ``horizontal''
(H) and ``vertical'' (V) with respect to the antenna frame -- simultaneously.
Due to the hardware layout of the correlators it is not yet possible to observe
all Stokes parameters. We collect linear polarization data on point sources via
the Earth rotation polarimetry, i.e. we monitor the fluxes in the H and V channels
as functions of parallactic angle $\psi$. The source polarization is derived from
the parameterization

\begin{equation}
q(\psi) = \frac{V-H}{V+H}(\psi) \equiv m_L\cos[2(\psi-\chi)]
\end{equation}

Here $m_L$ is the fraction of linear polarization (ranging from 0 to 1) and
$\chi$ is the polarization angle (ranging from 0$^{\circ}$ to 180$^{\circ}$
and counted from north to east). For details, please refer to \cite{key:trippe2010}.

\subsection{Kanata}

We performed the {\it V}-, {\it J}-, and {\it Ks}-band photometry and polarimetry
of \objectname{AO~0235+164} from 2008 August to 2008 October, using the TRISPEC instrument \citep{key:watanabe}
installed at the 1.5m Kanata telescope located at the Higashi-Hiroshima Observatory.
TRISPEC has a CCD and two InSb arrays, enabling photo-polarimetric
observations in one optical and two NIR bands simultaneously.  We obtained 21 photometric data points in the {\it V, J, Ks} bands.
A unit of the polarimetric observing sequence consisted of successive
exposures at 4 position angles of the half-wave
plates:  $0^{\circ},45^{\circ},22.5^{\circ},67.5^{\circ}.$
The data were reduced according to the standard procedure of CCD photometry.
We measured the magnitudes of objects with the aperture photometry technique.
We performed differential photometry with a comparison star taken in the
same frame of \objectname{AO~0235+164}. Its position is R.A.=02:38:32.31,
Dec=+16:35:59.7 (J2000)
and its magnitudes are {\it V} = 12.720, {\it J} = 11.248 and {\it Ks} =
10.711 \citep{key:gonzalez, key:cutri}.
The photometric data have been corrected for the Galactic extinction
of A({\it V}) = 1.473, A({\it J}) = 0.458 and A({\it Ks}) = 0.171, as explained
in Section~\ref{sec:extinction}.

We confirmed that the instrumental polarization was smaller than 0.1\%
in the {\it V} band using observations of unpolarized standard stars and
hence, we applied no correction for it. The zero point of the polarization
angle is corrected as standard system (measured from north to east) by
observing the polarized stars, HD19820 and HD25443 \citep{key:wolff}.

\subsection{SMARTS}
\objectname{AO~0235+164} was observed at the Cerro Tololo Inter-American
Observatory (CTIO) as part of a photometric monitoring campaign of
bright blazars with the Small and Moderate Aperture Research Telescope
System (SMARTS). The source was observed with the SMARTS 1.3m
telescope and ANDICAM instrument \citep{key:depoy}. ANDICAM is a
dual-channel imager with a dichroic linked to an optical CCD and an IR
imager, from which it is possible to obtain simultaneous data from 0.4
to 2.2 $\mu$m. Optical and near-infrared observations were taken in B,
V, R, J, and K bands.

Optical data were bias-subtracted, overscan-subtracted, and flat-fielded
using the CCDPROC task in IRAF. Infrared data were
sky-subtracted, flat-fielded, and dithered images were combined using
in-house IRAF scripts. The raw photometry of comparison stars in the
field of the blazar were calibrated using photometric zero-points that were
measured from 2008-2009 observations with ANDICAM of optical
\citep{key:landolt} and near-infrared \citep{key:persson} primary standards for
each filter, correcting for atmospheric extinction derived from all
the standards taken together.  The averages of the comparison stars
were used as a basis of differential photometry with respect to the
blazar for all observations. Errors were determined by calculating the
1$\sigma$ variation in the magnitude of the comparison stars.

Fluxes were computed using values for Galactic extinction from
\cite{key:schlegel1998}
and subtracting the nearby AGN `ELISA' as described
in \cite{key:raiteri2005}.
In addition, we accounted for the absorption of
the $z = 0.524$ system as outlined in \cite{key:junkkarinen}.

\subsection{Steward Observatory}
Optical spectropolarimetry and spectrophotometry of \objectname{AO~0235+164} during
fall 2008 was provided by the monitoring program being conducted at
Steward Observatory \citep{key:smith2009}.
This program utilizes the Steward Observatory CCD Spectropolarimeter (SPOL, \citealt{key:schmidt1992a})
at either the 2.3~m Bok telescope located on Kitt Peak, AZ, or the
1.54~m Kuiper telescope on Mt. Bigelow, AZ.  The publicly available data\footnote{\url{http://james.as.arizona.edu/~psmith/Fermi}}
include linear polarization and flux spectra (in 1st order) spanning
4000--7550~\AA.  General data-taking and reduction procedures used
for this project are described in detail in \cite{key:smith2003} and \cite{key:smith2009}.
For the monitoring of \objectname{AO~0235+164}, a $3''$ or $4''$-wide
slit was used for spectropolarimetry, depending on the observing conditions, and yielding
a spectral resolution of 20--25~\AA. An L-38 blocking filter was
inserted into the collimated beam for all observations to prevent
significant contamination from 2nd-order light until
well past 7600~\AA. Total exposure times of between 24 and 80 minutes were used
depending on the brightness of \objectname{AO~0235+164} and the sky/seeing conditions.
Usually, a high signal-to-noise-ratio measurement (${\rm S/N} > 100$) of the degree
of polarization (P) is determined from each observation by taking the
median linear, normalized Stokes parameters (q and u) in a 2000~\AA-wide
bin centered at 6000~\AA.  The reported values of P have been
corrected for statistical bias as in \cite{key:wardle}, but this
correction is typically not significant because of the high $S/N$ of the
binned data.  The position angle (theta) of the optical linear polarization
is calibrated by observing interstellar polarization standard stars \citep{key:schmidt1992b}.
Likewise, the flux spectra resulting from the spectropolarimetry
are calibrated using observations of spectrophotometric
standard stars \citep{key:massey}.  The flux spectra are corrected for
atmospheric extinction using the the standard extinction
curves given in \cite{key:baldwin} and \cite{key:stone}.
Flux information for \objectname{AO~0235+164} was obtained
through differential spectrophotometry of the blazar
and a nearby field star (``Star 4"; \citealt{key:gonzalez}).  The spectrophotometry employed slits
with widths of $7.6''$ or $12.7''$ to minimize seeing- and color-dependent
slit losses since the SPOL slit is left fixed in an east-west
orientation on the sky and is not aligned with the parallactic angle.
The wide-slit spectra of \objectname{AO~0235+164} and the comparison star were
convolved with standard filter transmission curves to determine
differential magnitudes and derive the apparent magnitude of the blazar
in the V and R bandpasses.  The spectrophotometric observations were much
shorter in duration (typically $< 5$ min) than the spectropolarimetry,
but of sufficient $S/N$ to be used to correct the much higher $S/N$ flux
spectra of \objectname{AO~0235+164} resulting from the spectropolarimetry for any
slit losses associated with the narrower slits used for those measurements.

\subsection{\textit{Swift} XRT and UVOT}

\objectname{AO~0235+164} was monitored as a result of an approved
target of opportunity (ToO) request by the \textit{Swift} satellite
\citep{key:gehrels2004} with weekly observations
of $\approx 1 - 2$ ks performed from 2008 September 2 to 2008 December
18 (Table~\ref{tab:swift}) with the X-ray Telescope (XRT; \citealt{key:burrows2005}) and
with the Ultraviolet/Optical Telescope (UVOT; \citealt{key:roming2005}).

The XRT data were reduced with the standard software ({\tt xrtpipeline
v0.12.4}) applying the default filtering and screening criteria (HEADAS
package, v6.9\footnote{\url{http://heasarc.gsfc.nasa.gov/lheasoft/}}).
We extracted the XRT light curve in the 0.3--10 keV
energy band using the software tool {\tt xrtgrblc}.  The source events were
extracted from circular regions centered on the source position. During the
outburst we excluded the inner 2 pixels of the source to avoid pile-up.
Exposure maps were used to account for the effects of vignetting,
point-spread function losses and the presence of hot pixels and hot
columns.

Since the source X-ray flux and spectrum are known to vary strongly,
co-adding individual XRT observations could be misleading.  We thus
extracted the XRT data from each individual pointing separately,
and fitted individual spectra using {\tt XSPEC}.  We rebinned the XRT data requiring at least 25 counts in each new energy bin.
As discussed in Sec. 3.1, we assumed the combined Galactic and $z=0.524$
absorption is adequately described by a column of
$2.8 \times 10^{21}$ cm$^{-2}$ at $z = 0$:  this
is in fact consistent with the spectral fit to the \textit{Swift} XRT data.
We determined the unabsorbed X-ray flux by performing the spectral fit
with fixed absorption, and then determining the incident flux by forcing
the absorption to be 0.  We include those fluxes in the 2 - 10 keV band
in the last column of Table \ref{tab:swift}.  We note that the source was detected at a
sufficiently good signal-to-noise (S/N) ratio to determine the spectrum unambiguously
only in the observations on MJD54711, MJD54758, and MJD54761.  In
other observations, we assumed a photon index of $\Gamma = 2$, consistent
with previous X-ray observations of this source in the low state, and
note that the error
resulting from such assumption on the inferred flux is comparable to
the statistical error quoted in the last column of Table~\ref{tab:swift}.

The UVOT photometry was done using the publicly available UVOT FTOOLS data
reduction suite and is based on the UVOT photometric system described in \cite{key:poole2008}
- but see also \cite{key:breeveld} for an updated calibration.
As discussed above, we adopted the corrections to the observed flux due
to the absorption by the Milky Way plus the intervening
galaxy at $z=0.524$ as outlined in Section~\ref{sec:extinction}.
The results of \textit{Swift} UVOT observations are presented in the Table~\ref{tab:uvot}.

\begin{table}
\begin{center}
\begin{tabular}{clll}
\hline
Date & exposure & Photon index & F$_{2-10 {\rm keV}}$ \\
(MJD-54000) & (sec) & &  $10^{-12}\ {\rm erg}\ {\rm cm}^{-2}\ {\rm s}^{-1}$ \\
\hline \hline
711.4976  & 6876  & $1.91\pm0.09$ & $3.1\pm 0.3$ \\
719.8695  & 1257  & $2$ (assumed) & $3.2\pm 0.4$ \\
737.9059  & 1448  & $2$ (assumed) & $3.7\pm 0.3$ \\
747.7476  & 2123  & $2$ (assumed) & $4.3\pm 0.3$ \\
758.7420  & 1133  & $2.44^{+0.07}_{-0.08}$ & $17.3\pm 1.4$ \\
761.7541  & 1181  & $2.60\pm0.08$ & $15.0\pm 1.2$ \\
781.0545  & 1144  & $2$ (assumed) & $2.8\pm 0.3$ \\
789.5603  & 1087  & $2$ (assumed) & $4.1\pm 0.4$ \\
803.7109  & 1175  & $2$ (assumed) & $4.5\pm 0.5$ \\
818.5249  & 1210  & $2$ (assumed) & $4.8\pm 0.4$ \\
\hline
\end{tabular}
\caption{The log of \textit{Swift} observations yielding good XRT data.  In all cases, the
spectrum was fitted with a power-law model absorbed by gas with Galactic abundances
with a column of $2.8 \times 10^{21}$ cm$^{-2}$ placed at $z = 0$:  such an
absorption form is only approximate, but it adequately fits
$ROSAT$ and $ASCA$ data, which in turn possess better signal-to-noise than
individual \textit{Swift} pointings (see text).  Since the quality of the data at MJD
54719, 54737, 54747, 54781, 54789, 54803, and 54818 have too low a $S/N$ for reliable
determination of spectrum, we assumed a photon index of 2 for those pointings.  }
\label{tab:swift}
\end{center}
\end{table}

\begin{sidewaystable}
\scriptsize
\begin{center}
\begin{tabular}{l|ll|ll|ll|ll|ll|ll}
\hline
Date & $M(v)$ & F(v) & $M(b)$ & $F(b)$ & $M(u)$ & $F(u)$ & $M(w1)$ & $F(w1)$ & $M(m2)$ & $F(m2)$ & $M(w2)$ & $F(w2)$ \\
\hline
\hline
711.50  & $16.95\pm0.05$ & 2.32  & $17.93\pm0.04$ & 1.49  & $18.07\pm0.06$ & 1.04  & $18.16\pm0.06$ & 0.49  & $18.57\pm0.07$ & 0.43  & $18.93\pm0.06$ & 0.28  \\
719.87  & $16.79\pm0.09$ & 2.69  & $17.81\pm0.08$ & 1.65  & $17.75\pm0.11$ & 1.39  & $17.99\pm0.11$ & 0.57  & $18.23\pm0.16$ & 0.59  & $18.47\pm0.10$ & 0.42  \\
729.65  & $16.17\pm0.07$ & 4.76  & $17.03\pm0.06$ & 3.41  & $17.13\pm0.08$ & 2.46  & $17.18\pm0.08$ & 1.21  & $17.46\pm0.11$ & 1.20  & $17.85\pm0.08$ & 0.75  \\
737.91  & $16.29\pm0.06$ & 4.26  & $17.11\pm0.05$ & 3.17  & $17.11\pm0.07$ & 2.51  & $17.47\pm0.08$ & 0.93  & $17.62\pm0.11$ & 1.03  & $18.04\pm0.08$ & 0.63  \\
740.65  & $15.89\pm0.06$ & 6.18  & $16.80\pm0.05$ & 4.20  & $17.00\pm0.08$ & 2.77  & $17.17\pm0.08$ & 1.22  & $17.37\pm0.10$ & 1.30  & $17.78\pm0.08$ & 0.80  \\
747.75  & $16.17\pm0.05$ & 4.76  & $17.01\pm0.04$ & 3.46  & $17.14\pm0.06$ & 2.45  & $17.36\pm0.06$ & 1.03  & $17.43\pm0.07$ & 1.23  & $17.94\pm0.06$ & 0.69  \\
758.74  & $15.93\pm0.06$ & 5.95  & $16.86\pm0.05$ & 3.96  & $16.85\pm0.07$ & 3.18  & $16.83\pm0.07$ & 1.67  & $17.10\pm0.09$ & 1.66  & $17.61\pm0.07$ & 0.93  \\
761.75  & $16.05\pm0.06$ & 5.33  & $16.92\pm0.05$ & 3.76  & $16.99\pm0.07$ & 2.79  & $17.09\pm0.08$ & 1.31  & $17.29\pm0.09$ & 1.40  & $17.55\pm0.07$ & 0.99  \\
768.80  & \nodata & \nodata & $17.06\pm0.07$ & 3.30  & $17.08\pm0.10$ & 2.58  & $16.96\pm0.07$ & 1.47  & \nodata & \nodata & $17.45\pm0.12$ & 1.08  \\
780.31  & $17.07\pm0.16$ & 2.07  & $17.99\pm0.15$ & 1.40  & $17.91\pm0.19$ & 1.2  & $18.14\pm0.17$ & 0.50  & $18.17\pm0.28$ & 0.62  & $18.46\pm0.15$ & 0.43  \\
781.06  & $17.31\pm0.13$ & 1.66  & $17.85\pm0.09$ & 1.59  & $17.97\pm0.14$ & 1.13  & $18.08\pm0.13$ & 0.53  & $18.39\pm0.17$ & 0.51  & $18.63\pm0.11$ & 0.37  \\
789.56  & $16.63\pm0.08$ & 3.11  & $17.50\pm0.07$ & 2.21  & $17.46\pm0.10$ & 1.81  & $17.81\pm0.11$ & 0.68  & $17.98\pm0.16$ & 0.74  & $18.25\pm0.09$ & 0.52  \\
790.83  & $16.60\pm0.11$ & 3.20  & $17.40\pm0.09$ & 2.41  & $17.80\pm0.16$ & 1.32  & $17.73\pm0.14$ & 0.73  & $18.04\pm0.18$ & 0.70  & $18.34\pm0.13$ & 0.48  \\
803.71  & $17.66\pm0.16$ & 1.21  & $18.77\pm0.18$ & 0.69  & $18.63\pm0.21$ & 0.62  & $18.79\pm0.20$ & 0.27  & $18.51\pm0.17$ & 0.46  & $19.19\pm0.16$ & 0.22  \\
813.64  & $18.10\pm0.24$ & 0.81  & $18.87\pm0.21$ & 0.63  & $18.97\pm0.31$ & 0.45  & $18.96\pm0.24$ & 0.23  & $19.00\pm0.28$ & 0.29  & $19.00\pm0.16$ & 0.26  \\
818.52  & $17.43\pm0.15$ & 1.49  & $18.34\pm0.15$ & 1.01  & $18.87\pm0.32$ & 0.50  & $18.63\pm0.20$ & 0.32  & $19.07\pm0.23$ & 0.27  & $19.17\pm0.17$ & 0.22  \\
\hline
\end{tabular}
\caption{Results of \textit{Swift} UVOT observations of AO~0235+164. The data are listed for six \textit{Swift} UVOT filters.  Each pair of columns corresponds to the observed, uncorrected magnitude $M$ (left entry) and corrected flux density $F$, in
units of milliJansky (right entry).  To correct
for absorption in the Milky Way plus that at $z = 0.524$, we used the following
values of absorption $A_{\lambda}$ for the respective UVOT filters:
W2: 2.87; M2: 2.94; W1: 2.52; U: 2.71; B: 1.84; V: 1.46.
%W2: 3.637; M2: 3.288; W1: 2.777; U: 2.519; B: 1.904; V: 1.473.
We calculate the
incident flux $F_{\rm inc, \lambda}$ in the band corresponding to $\lambda$ from the observed (absorbed)
flux $F_{\rm abs, \lambda}$ via $F_{\rm abs, \lambda} / F_{\rm inc, \lambda} = 10^{A_{\lambda}/2.5}$.
The same corrections are applied to ground-based optical data.}
\label{tab:uvot}
\end{center}
\end{sidewaystable}

\subsection{RXTE}
As part of our campaign, 30 observations of \objectname{AO~0235+164} were obtained with $Rossi$ X-ray Timing
Explorer ($RXTE$) between 2008 October 18 and 2008 December 27.  We analyzed
the data from the Proportional Counter Array (PCA) following standard
procedures. We selected only data from PCU2, the best calibrated module
and the only one which is always turned on. The data were screened in the
following way: source elevation above the horizon $>$ 10$^\circ$, pointing
offset smaller than 0.02$^\circ$, at least 30 minutes away from a South Atlantic Anomaly passage
and electron contamination smaller than 0.1. This resulted in a total exposure
of 192.3 ks. Single net PCA exposures range from 2.1 ks to 14.1 ks. Background
was estimated with standard procedures and the detector response matrices
extracted with the $RXTE$ tools (command PCARSO v. 10.1).

For the spectral analysis
the fitting procedure was done with the
{\tt XSPEC} software package. The spectra from the channels corresponding to nominal energies of
2.6 to 10.5 keV are adequately fitted
by a single power law model, absorbed by a fixed column of $2.8 \times 10^{21}$
cm$^{-2}$ at $z = 0$ as determined by the ROSAT and ASCA - in an analogous manner
to the spectral fitting performed to the \textit{Swift} XRT data above.
The parameters of the fits are reported in Table~\ref{tab:rxte};  again,
the last column reports the unabsorbed X-ray flux.

\begin{table}[htdp]
\begin{center}
\begin{tabular}{c|c|c|c|c|c}
\hline
Date & MJD-54000 & exposure (s) & photon index $\Gamma$ &  $\chi ^2_r$ / d.o.f. & $F_{2-10}$  \\
\hline
\hline
 18/10/2008 18:56 & 757.805 & 2688  & $2.46 \pm 0.13$    &   0.60/9 &     $2.00 \pm 0.08  $\\
 19/10/2008 13:35 & 758.684 & 12416 & $2.55 \pm 0.06$    &   0.41/9 &     $2.09 \pm 0.04  $\\
 20/10/2008 13:09 & 759.566 & 3024  & $2.73 \pm 0.14$    &   0.62/9 &     $1.79 \pm 0.07  $\\
 21/10/2008 14:16 & 760.647 & 6384  & $2.56 \pm 0.10$    &   0.38/9 &     $1.70 \pm 0.05  $\\
 22/10/2008 13:53 & 761.597 & 3104  & $2.34 \pm 0.12$    &   0.56/9 &     $1.87 \pm 0.07  $\\
 23/10/2008 18:08 & 762.773 & 2976  & $2.47 \pm 0.12$    &   0.58/9 &     $2.03 \pm 0.07  $\\
 25/10/2008 14:08 & 764.704 & 12656 & $2.57 \pm 0.07$    &   0.43/9 &     $1.77 \pm 0.04  $\\
 26/10/2008 15:13 & 765.653 & 3136  & $2.71 \pm 0.14$    &   0.57/9 &     $1.68 \pm 0.08  $\\
 27/10/2008 13:09 & 766.567 & 3200  & $2.35 \pm 0.16$    &   0.83/9 &     $1.41 \pm 0.07  $\\
 28/10/2008 19:01 & 767.806 & 2320  & $2.71 \pm 0.21$    &   0.34/9 &     $1.26 \pm 0.08  $\\
 31/10/2008 14:32 & 770.624 & 3152  & $2.24 \pm 0.27$    &   0.44/9 &     $0.71 \pm 0.06  $\\
 02/11/2008 17:03 & 772.850 & 14448 & $2.26 \pm 0.18$    &   0.29/9 &     $0.53 \pm 0.03  $\\
 03/11/2008 13:07 & 773.566 & 3200  & $2.46 \pm 0.40$    &   0.55/9 &     $0.54 \pm 0.07  $\\
 04/11/2008 19:11 & 774.880 & 9072  & $2.72 \pm 0.36$    &   0.34/9 &     $0.37 \pm 0.04  $\\
 05/11/2008 18:34 & 775.858 & 9680  & $2.12 \pm 0.31$    &   0.48/9 &     $0.35 \pm 0.04  $\\
 06/11/2008 18:07 & 776.774 & 3200  & $1.91 \pm 0.55$    &   0.59/9 &     $0.32 \pm 0.06  $\\
 07/11/2008 14:32 & 777.723 & 12992 & $2.50 \pm 0.38$    &   0.45/9 &     $0.28 \pm 0.03  $\\
 10/11/2008 16:07 & 780.696 & 2880  & $3.53 \pm 0.79$    &   0.79/9 &     $0.38 \pm 0.09  $\\
 11/11/2008 11:11 & 781.486 & 3264  & $2.37 \pm 0.61$    &   1.40/9 &     $0.27 \pm 0.06  $\\
 13/11/2008 11:53 & 783.613 & 13168 & $2.37 \pm 0.38$    &   0.39/9 &     $0.25 \pm 0.03  $\\
 14/11/2008 12:56 & 784.625 & 9888  & $2.39 \pm 0.36$    &   0.59/9 &     $0.30 \pm 0.04  $\\
 15/11/2008 11:00 & 785.543 & 9904  & $2.37 \pm 0.36$    &   0.63/9 &     $0.31 \pm 0.04  $\\
 17/11/2008 10:09 & 787.474 & 6336  & $2.50 \pm 0.59$    &   0.46/9 &     $0.24 \pm 0.05  $\\
 19/11/2008 09:10 & 789.429 & 5568  & $2.94 \pm 0.44$    &   0.81/9 &     $0.39 \pm 0.05  $\\
 20/11/2008 07:10 & 790.411 & 12192 & $1.99 \pm 0.36$    &   0.34/9 &     $0.26 \pm 0.03  $\\
 21/11/2008 16:48 & 791.366 & 9744  & $2.92 \pm 0.63$    &   0.39/9 &     $0.23 \pm 0.04  $\\
 23/11/2008 10:33 & 793.492 & 6432  & $2.08 \pm 0.68$    &   0.42/9 &     $0.21 \pm 0.04  $\\
 27/12/2008 14:06 & 827.672 & 9664  & $2.69 \pm 0.50$    &   0.40/9 &     $0.25 \pm 0.04  $\\
 \hline
\end{tabular}
\caption{Best-fit parameters for the PCA data of each RXTE  observation
with the absorption fixed at the value measured by ROSAT + ASCA, with the  column
of $N_H= 28 \times 10^{20}$ cm$^{-2}$ with Galactic abundances.
Description of columns: (1) and (2) Observing date, (3) Exposure (s),
(4) photon index and error,
(5) reduced $\chi^{2}$ and no. of degrees of freedom,
(6) Flux in the 2-10 keV band, in units of $10^{-11}$ erg cm$^{-2}$ s$^{-1}$.}
\label{tab:rxte}
\end{center}
\end{table}

\section{Variability of the Source}
\label{sec:mw_lightcurve}

\subsection{Multi-wavelength Light Curve}

\begin{figure}[htbp]
\begin{center}
\includegraphics[width=1.0\textwidth]{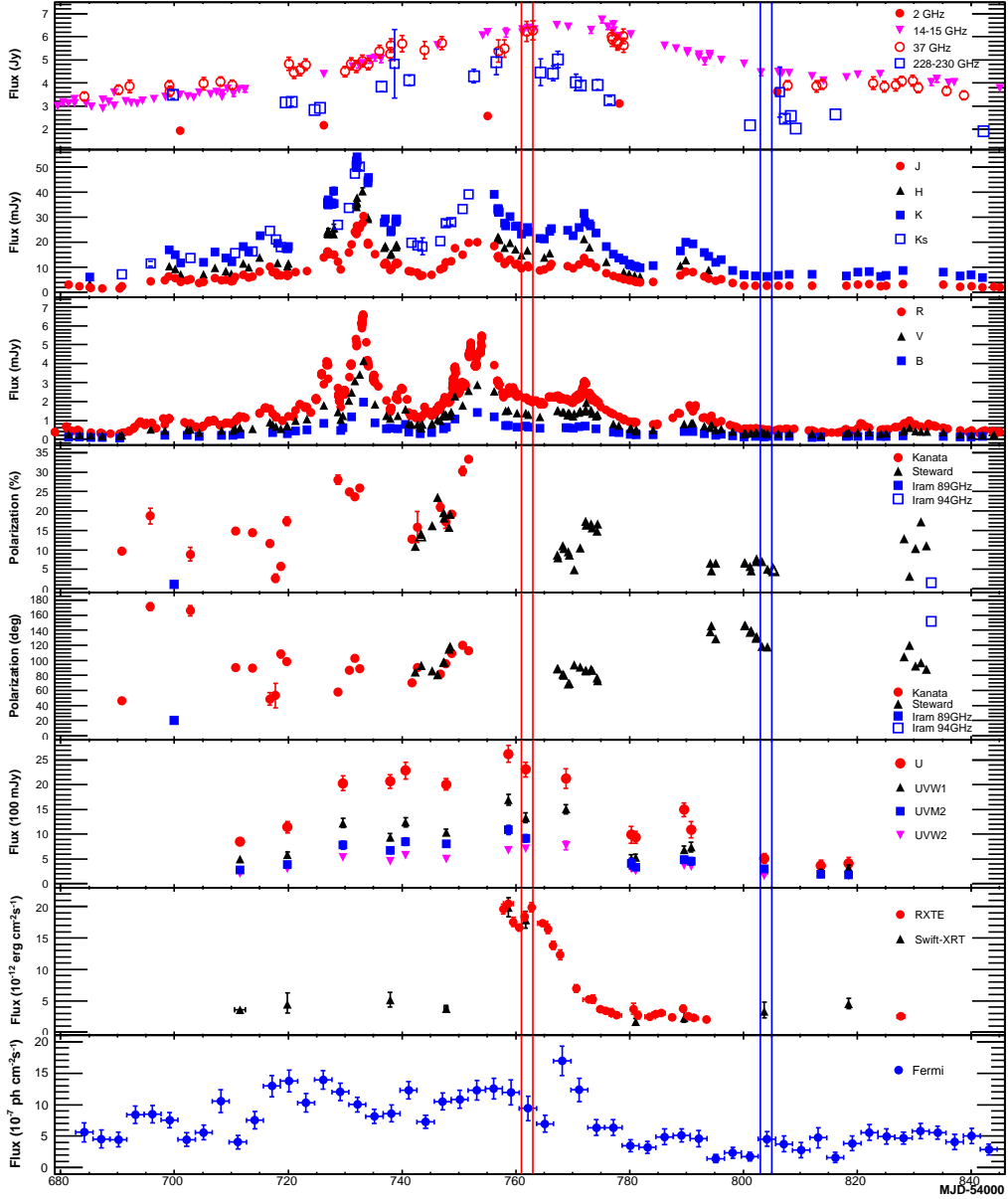}
\caption{AO~0235+164 light curve from 2008 August 4 to 2009 February 4
in different energy ranges. From the top to the bottom: radio, near IR,
optical, UV, X-rays and $\gamma$-rays above 100~MeV. Panels 4 and 5
from the top report the polarization data from the Kanata optical
observatory and IRAM radio telescope. Two double vertical lines
mark the epochs for which we extracted the SEDs modeled in Section \ref{sec:modeling}.}
\label{fig:LightCurve}
\end{center}
\end{figure}

In this section we present the results of the multi-wavelength observations conducted on
\objectname{AO~0235+164} from 2008 August to 2009 January.
Figure~\ref{fig:LightCurve} shows
the multi-wavelength data available. From the top to the bottom are:
radio, near-infrared, optical, polarization degree ($\%$) and
polarization angle (deg), UV, X-ray and $\gamma$-ray from 100 MeV
to 100 GeV data are grouped together.

\begin{figure}[htbp]
\begin{center}
\includegraphics[width=0.8\textwidth]{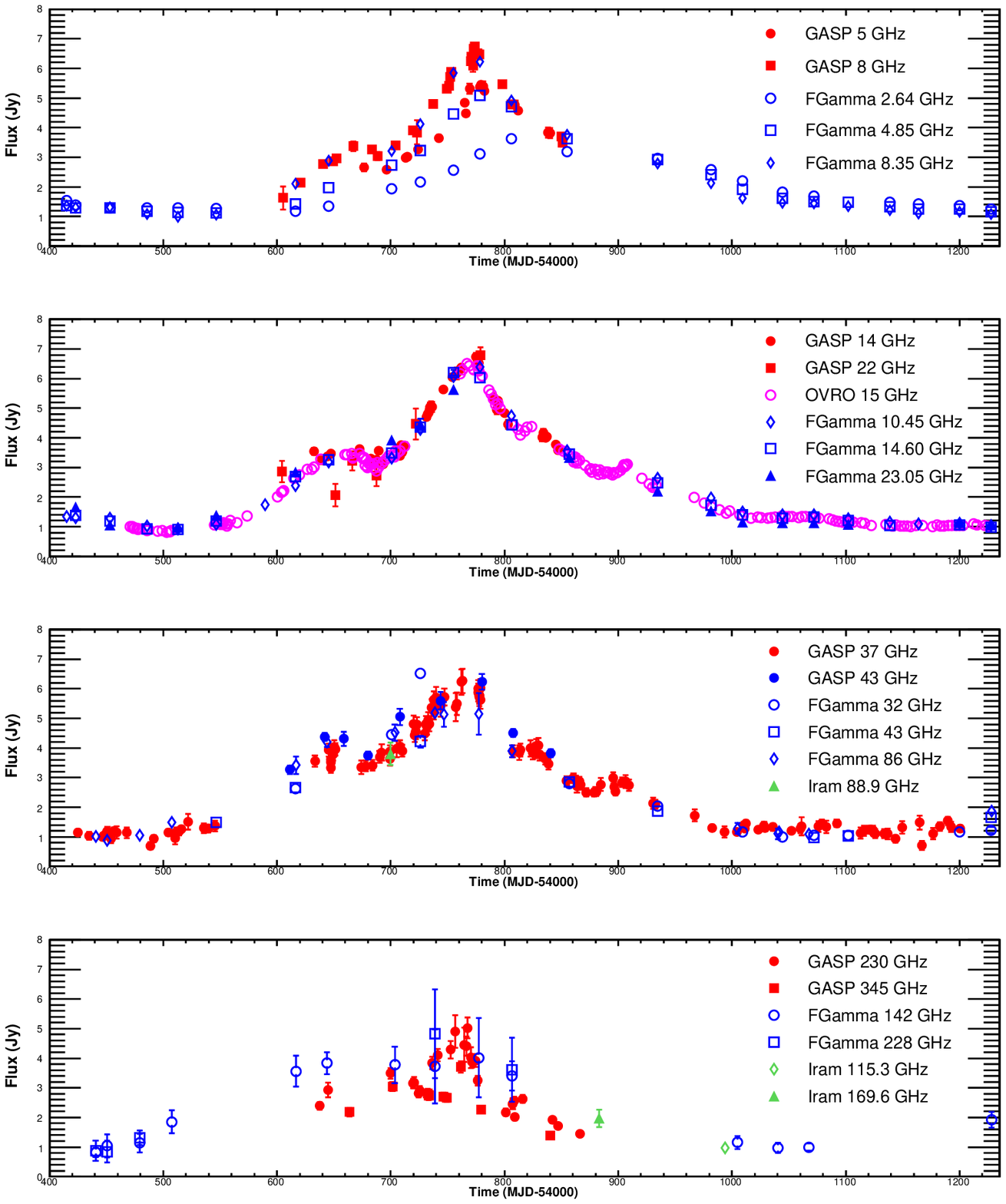}
\caption{Radio,  mm and sub-mm light
curves of AO~0235+164 from mid 2007 until June 2010 (MJD 54400-55230). In these energy bands the
flux began to increase around the middle of 2007 (MJD 54500), months before the
start of the optical and higher energy activity of the source.}
\label{fig:extendedradioLC}
\end{center}
\end{figure}

The optical behavior is the best sampled among all. Two main flare
peaks are visible in the period around 2008 October (MJD 54730-54750), and they are
surrounded by other smaller peaks. The radio data show that the flux started to increase smoothly starting around the middle of 2007 (MJD 54500, which is apparent
in Figure~\ref{fig:extendedradioLC}), reaching its
maximum during the optical flare activity and slowly
decreasing when the source returned to a low flux state
in near IR, optical, X-ray and $\gamma$-ray bands.
The near infrared data show the same temporal trend as the
optical bands. The UV data from \textit{Swift} UVOT do not show the level of
activity seen in the optical band.  The X-ray data from
\textit{Swift} XRT and RXTE present a very pronounced
peak clearly delayed with respect to the optical activity.

The \textit{Fermi} light curve, as already discussed in section
\ref{sec:gammalc} shows a broad high-state period followed
by a final narrow peak succeeding the X-ray peak before
getting to the low-flux state. Since then (up to the time of submitting this paper in autumn 2011), the source has been in a very quiet state.

Figure~\ref{fig:extendedradioLC} shows the light curves constructed from
the radio, mm and sub-mm data in an extended time interval, from 2007
to 2010 June (MJD 54400-55230). In the lowest-energy-band, the increasing
trend of the flux started months before the increased level
of activity seen in the optical and higher
energy bands.  After the period of the increased radio/mm flux
associated with the optical flaring activity, the source
enters a period of gradually declining flux.

\subsection{Cross-correlation studies and time delays}

\begin{figure}[htbp]
\begin{center}
\includegraphics[width=0.61\textwidth]{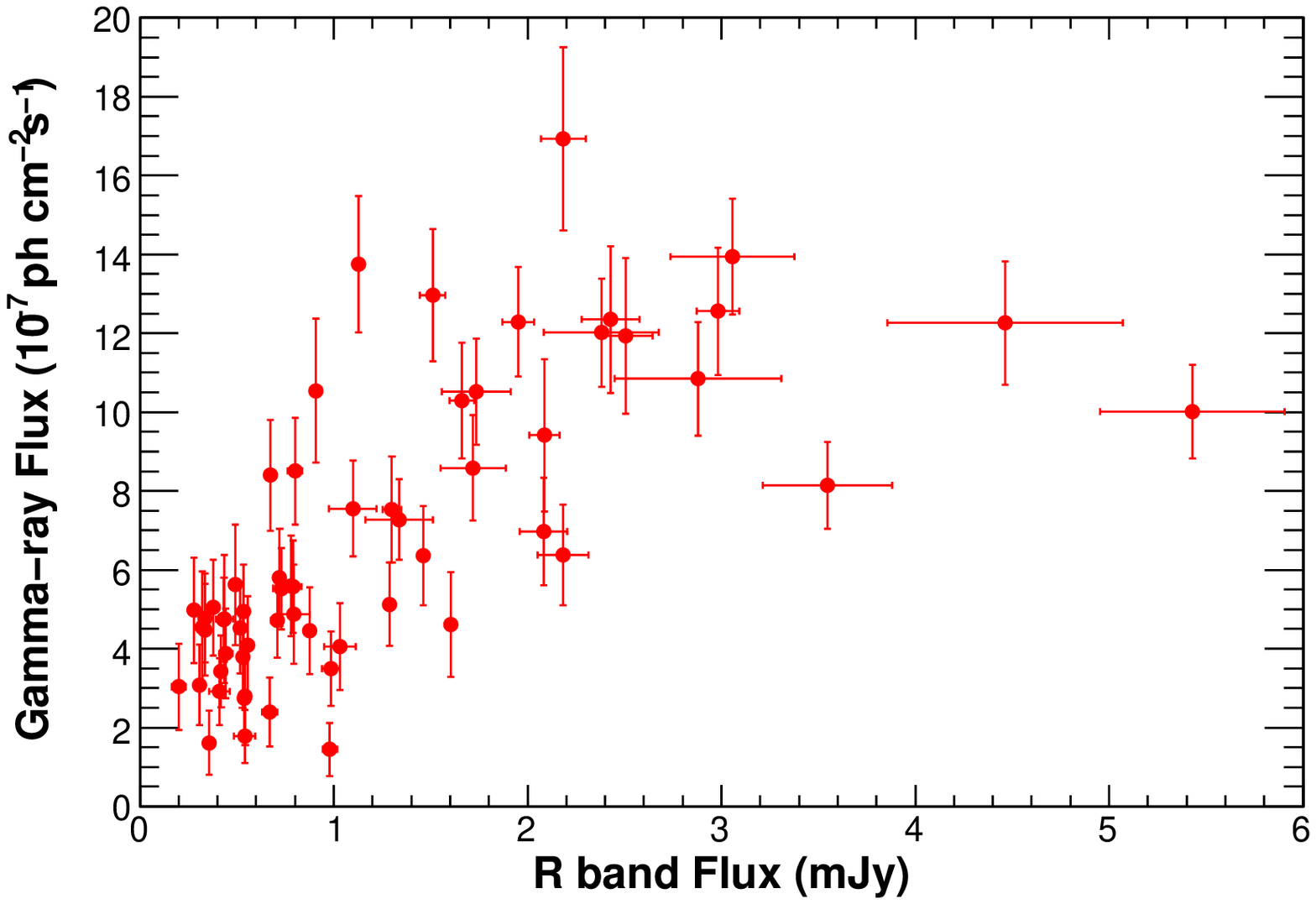}
\includegraphics[width=0.61\textwidth]{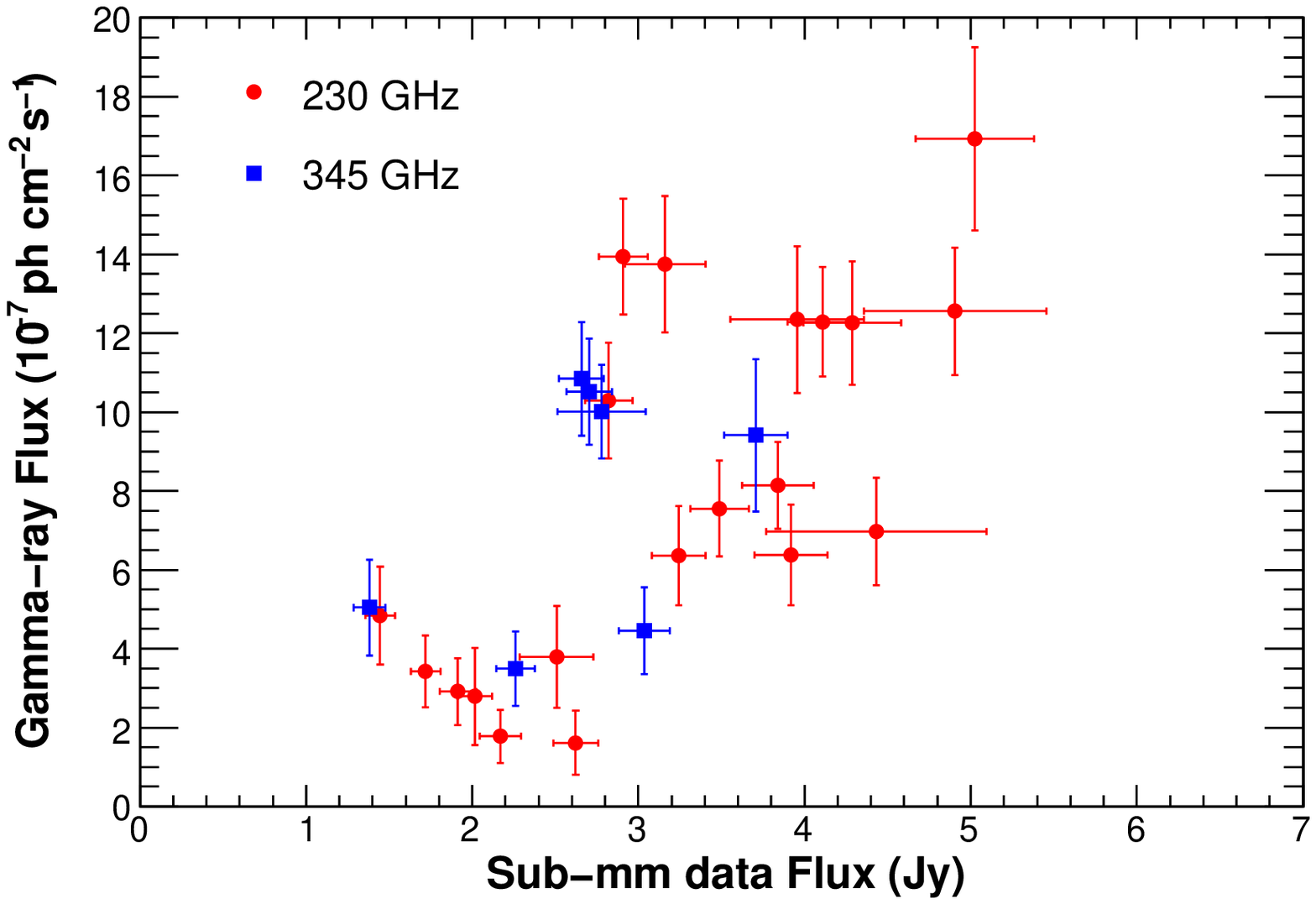}
\caption{Top: Plot of the $\gamma$-ray flux vs R band flux; both fluxes are averaged
in 3-days time intervals.
The data suggest that $\gamma$-ray and optical fluxes follow each other, but the correlation is small
with $\gamma$-ray flux reaching a plateau at the level $\sim 1.2 \times 10^{-6}$ when
the optical flux reaches $\sim 3$ mJy, but not increasing beyond $\sim 1.2 \times 10^{-6}$
when the optical flux increases to $\sim 5$ mJy.
Bottom: Similar plot of $\gamma$-ray flux vs high-frequency radio-band flux; both fluxes are averaged
in 3-day time intervals. Likewise, there is a general trend of increase in
both bands, but the correlation is small.}
\label{fig:rVSgamma}
\end{center}
\end{figure}

We searched for correlations of variability between different bands, with the goal
to understand the relationship between the fluxes of
\objectname{AO~0235+164} at different energies.
The cross-correlation studies between the optical
R band and $\gamma$-ray fluxes are illustrated in the
the top panel of Figure~\ref{fig:rVSgamma}.
Those data have a Spearman correlation
coefficient\footnote{Wessa, P. (2011), Free Statistics Software, Office
for Research Development and Education, version
1.1.23-r7, \url{http://www.wessa.net/}} of 0.75 \citep{key:spearman}.
The relations between the $\gamma$-ray and 230~GHz and 345~GHz fluxes have also been evaluated and the results are shown in Figure~\ref{fig:rVSgamma}
in the bottom panel: the Spearman correlation coefficient between $\gamma$-ray
fluxes and 230~GHz
data is 0.70 showing that there exists a correlation between the two data sets.
On the other hand, the sampling at 345~GHz is poor, with only a few data points at that frequency and the evaluation of a correlation has not been performed.
In all cases, no correlation is found at 90$\%$ confidence level.
We note here that \cite{key:agudo11}, using data collected for this object
over a longer time span than that covered by our observations, performed
a light-curve correlation analysis following the method described by \cite{key:agudo11a}. They found that these bands are correlated at a 99.7\% confidence
level. In our case, no correlation is found at 90\% confidence level:  it is very likely
that the stronger correlation of signals derived by \cite{key:agudo11}
is caused by their use of a significantly longer time span,
amounting to roughly 8 years.  All this suggests that the variability
of the source in the $\gamma$-ray and radio-to-mm regimes on long time
scales is correlated, but the situation on shorter time scales is less clear.

\begin{figure}[htbp]
\begin{center}
\includegraphics[width=0.57\textwidth]{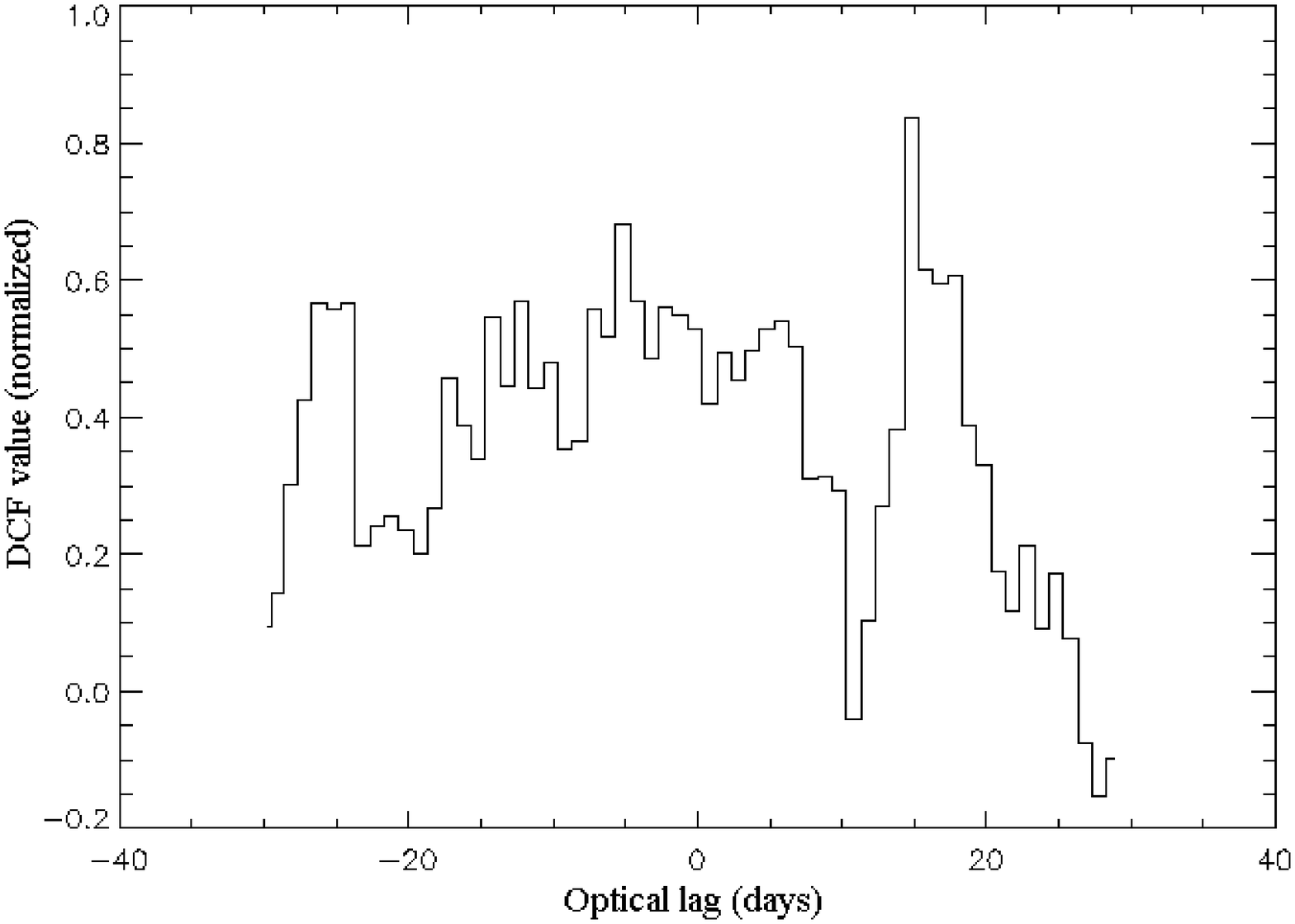}
\caption{Discrete Correlation Function (DCF) calculated between optical
R band and \textit{Fermi}-LAT $\gamma$-ray data binned over 1-day intervals.
Positive values correspond to $\gamma$-rays leading the optical signal.}
\label{fig:dcf}
\end{center}
\end{figure}

Since the time
series in the optical R band and $\gamma$-ray are the best sampled in this study, it was possible to calculate lags/leads between those bands.
To this end, we calculated the discrete correlation function (DCF, \citealt{key:dcf}).
We binned the data sets in
order to smooth the intra-day features in the optical light curves,
and to obtain similar sampling in the $\gamma$-ray band.
We tried several bin sizes from 1 to 7 days to check how
sensitive the results are to this smoothing procedure.
The DCF from the optical and $\gamma$-ray data
do not show significant peaks on short time scales (1~day) meaning no
optical-$\gamma$ correlation is detected over the observing period.
Figure~\ref{fig:dcf} shows the result of this DCF analysis
when the light curves are binned over 1 day.  A peak
can be seen at 15 days (with optical lagging $\gamma$-rays), however
the significance is modest. \cite{key:agudo11} found that, for a similar
period, the optical flux lags the $\gamma$-rays by $\sim 10$ days,
but their DCF peak is much broader, and could be interpreted as being consistent with no lag.

\subsection{Time dependence of optical polarization}
\label{sec:polarization}

As illustrated in Figure~\ref{fig:LightCurve},
the polarization degree and angle
are highly variable;  the former correlates with the optical flux and
at the two largest flux peaks reaches values ~25\% and ~35\%, respectively.
This correlation was studied over a longer period of time
(from 2008 August 12 to 2009 February 18) by \cite{key:sasada11},
and in the past, during the outburst of 2006 December, by
\cite{key:hagen08}. A trend of the stabilization
of the polarization angle during flares is seen both
in 2006 and  2008, but around different values, with electric vector polarization angle (EVPA) at
$\sim -30^{\circ}$ and $\sim 100^{\circ}$ respectively.
Comparing EVPA  with the position angle of
the parsec-scale jets, \cite{key:hagen08} found that there is a trend of
their alignment during high states. However, since the parsec-scale jet
in \objectname{AO~0235+164} shows large changes of direction with time
\citep{key:jones84,key:chu96,key:jorstad,key:piner06}
and the jet direction
to which  EVPA was compared  was inferred from the VLBI maps taken in different
 epochs, the claimed alignment could be accidental. Indeed, comparison of
EVPA during flux peaks in 2008 with the direction of the jet determined during
the same epoch by VLBI observations does not confirm such
an alignment \citep{key:agudo11}.
On the contrary,  both angles are	
oriented perpendicular rather than parallel to each other, albeit with a large scatter, with EVPA at
optical flux peaks $\sim 100^{\circ}$ vs. $\chi_{jet} \sim 0^{\circ}$.
This implies a parallel orientation of the magnetic fields to the jet and may
indicate production of flares in a reconfinement shock \citep{key:Nalewajko09}.

\section{Broad-band Spectral Energy Distribution}
\label{sec:sed}

Our unprecedented time sampling of \objectname{AO~0235+164}
in several spectral bands allows us to extract
accurate instantaneous SEDs, which are needed to correctly
interpret the broad-band emission of the source.
We reiterate that in order to build the intrinsic SED and correctly convert the observed magnitudes to de-absorbed fluxes, extinction must be taken into account, including both Galactic extinction, and that due to the $z = 0.524$ system.
We discussed in Section \ref{sec:extinction} how the IR, optical, UV data and soft X-rays are affected by the absorption due to both systems.
Following the prescriptions from \cite{key:junkkarinen},
we assume that the soft X-ray absorption is adequately described by a column density of $2.8 \times 10^{21}\;{\rm cm^{-2}}$ at $z = 0$. However, for the far-UV data, we use a modified extinction model based on the work of \cite{key:pei}.

Figure~\ref{fig:sed_overall} shows the broad-band SEDs obtained by
plotting simultaneous radio, NIR, optical, UV, X-ray data in the following
two 2-day epochs:
\begin{enumerate}
\item
MJD 54761--54763 (shown in red), corresponding to the maximum of the X-ray flare,
coincident with a highly variable near-IR/optical/UV state
and a high $\gamma$-ray state.
\item
MJD 54803--54805 (shown in blue), corresponding to a low state in all bands,
following the high-activity period.
\end{enumerate}
The plotted data points were extracted from the larger datasets as follows:
\begin{itemize}
\item[-] Radio data:
most of the data points are simultaneous measurements.
Although the sampling at some particular wavelengths is poor,
the available radio light curves show very smooth and slow trends,
thus we have also plotted interpolated values based on the extended data
set of about 10 days in length, centered on the main observation period.
\item[-]
Near-IR/optical/UV: all the data shown are simultaneous measurements
made by GASP-WEBT, SMARTS and \textit{Swift} UVOT telescopes.
\item[-]
X-ray data: in the first epoch (MJD 54761--54763), we present
the \textit{Swift} XRT observation with a butterfly plot.
In the second epoch (MJD 54803--54805), the $S/N$ for the
\textit{Swift} XRT measurement is too low to allow a good spectral
representation and only the flux upper limit is reported.
\item[-]
$\gamma$-ray data: the $\gamma$-ray spectra have been built
following the analysis procedure described in Section \ref{sec:gammaspectrum}.
Since the time intervals chosen to build the broad-band SED are too short
to allow a good reconstruction of the $\gamma$-ray spectrum, longer periods
have been used.
The first time interval, MJD 54750--54770,
corresponds to a period of high $\gamma$-ray emission,
which includes the X-ray flare period.
The high state is followed by a lower emission state whose
spectrum is averaged in the interval MJD 54780--54840.
\end{itemize}

The overall SED, plotted in Figure~\ref{fig:sed_overall},
appears quite similar to that measured for other blazars.
There is one marked difference:
at least for the first period, the X-ray spectrum is soft,
yet it is not located on the extrapolation of the optical/UV spectrum.
%It is reasonable to ask wheter that the extinction correction derived by \cite{key:junkkarinen} and used here could be too low, as may be indicated by a sharp hardening in the UV band that is independent of the luminosity in this band.
For this to be the case, extinction
would have to be significantly greater, with the error at the level of
at least 50\%, which we consider unlikely.
%but not completely out of the question, especially given the two-component nature of the intervening absorber.
Assuming that we adopted the correct extinction, the broad-band SED does show a distinct
feature in the soft X-ray band, separate from the two broad peaks forming
the SED in most blazars, and we discuss its origin below.

\section{Modeling of the Broad-Band Spectrum}
\label{sec:modeling}

\objectname{AO~0235+164}, like many other
luminous, low-frequency-peaked BL Lac objects,
shows broad emission lines \citep{key:cohen87,key:nilsson96,key:raiteri07}.
Using the emission-line spectrum reported
in \cite{key:raiteri07}, correcting the line flux for extinction, and assuming
that the contribution of the lines measured by them to the total luminosity of broad emission
lines (BEL) is the same as in the composite spectrum of quasars
\citep{key:francis91}, we find
$L_{\rm BEL} \sim 4 \times 10^{44}\;{\rm erg\, s^{-1}}$. For the typical
covering factor of the broad-line region (BLR) $\xi_{\rm BEL} \sim 0.1$
this implies a luminosity of the accretion disk of $L_d \sim 4 \times 10^{45}\;{\rm erg\,s^{-1}}$.
With such a high accretion luminosity, if observed directly, i.e.
without being overshone by the jet nonthermal radiation, \objectname{AO~0235+164}
would satisfy a formal criterion to be classified as a quasar \citep[see also][]{key:MBP93}.
This means that, according the the AGN unification models, it should possess a typical dusty torus, a strong source of thermal infrared radiation (IR) with a typical covering factor of $\xi_{\rm IR}\sim 0.1$ \citep[e.g.,][]{key:haas}.
Recent mid-IR interferometric observations for a sample of nearby AGN show that
such tori can extend beyond 10 pc from the central black hole \citep{key:tristan2011}.
The mass of the black hole (BH) in this object is likely to be
in the range $M_{BH} \sim 2-6 \times 10^8\;M_{\odot}$
\citep{key:liu06,key:raiteri07,key:wu10}, which implies the
Eddington ratio  $L_d/L_{\rm EDD} \ge 0.1$.

In order to determine which process dominates the high-energy emission,
whether it is ERC or SSC, one can estimate their luminosity ratio as
$L_{\rm ERC}/L_{\rm SSC}\simeq u_{\rm ext}'/u_{\rm syn}'$, where $u_{\rm ext}'$ is the co-moving energy density of the external radiation, which depending on the source location could be dominated either by $u_{\rm BEL}'$ or $u_{\rm IR}'$, and $u_{\rm syn}'$ is the co-moving energy density of the synchrotron radiation. These energy densities scale like $u_{\rm BEL(IR)}'\simeq\Gamma_{\rm j}^2u_{\rm BEL(IR)}\simeq\Gamma_{\rm j}^2\xi_{\rm BEL(IR)}L_{\rm d}/(4\pi r_{\rm BEL(IR)}^2c)$ for $r\le r_{\rm BEL(IR)}$, respectively, and $u_{\rm syn}' \simeq L_{\rm syn}/ (4 \pi R^2 {\cal D}^4c)$, where $r_{\rm BEL}\sim 0.1(L_{\rm d,46})^{1/2}\;{\rm pc}$ is the characteristic radius of the broad-line region, $r_{\rm IR}\sim 2.5(L_{\rm d,46})^{1/2}\;{\rm pc}$ is the inner radius of the dusty torus, $R$ is the emitting zone radius related to its distance by $r=R \Gamma_{\rm j}$, $\Gamma_{\rm j}=(1-\beta_{\rm j}^2)^{-1/2}$ is the jet Lorentz factor and $\beta_{\rm j}$ is the jet velocity in units of $c$ \citep{key:sikora09}.
Considering the emitting zone located at either characteristic radius, i.e. $r\simeq r_{\rm BEL(IR)}$, and neglecting the distinction
between the Doppler factor ${\cal D}$ and the Lorentz
factor $\Gamma_{\rm j}$, we obtain $L_{\rm ERC}/L_{\rm SSC}\simeq \xi_{\rm BEL(IR)}\Gamma_{\rm j}^4(L_{\rm d}/L_{\rm syn})$.
In the case of \objectname{AO~0235+164},
we observe $L_{\rm d}/L_{\rm syn}\sim 0.01$ and thus $L_{\rm ERC}/L_{\rm SSC}\simeq 160(\xi_{\rm BEL(IR)}/0.1)(\Gamma_{\rm j}/20)^4$.
Hence, even for a moderate bulk Lorentz factor,
in order for the SSC component to dominate the ERC component,
one requires covering factors 2 orders of magnitude lower than typically assumed in quasars.

\begin{figure}
  \includegraphics[width=\textwidth]{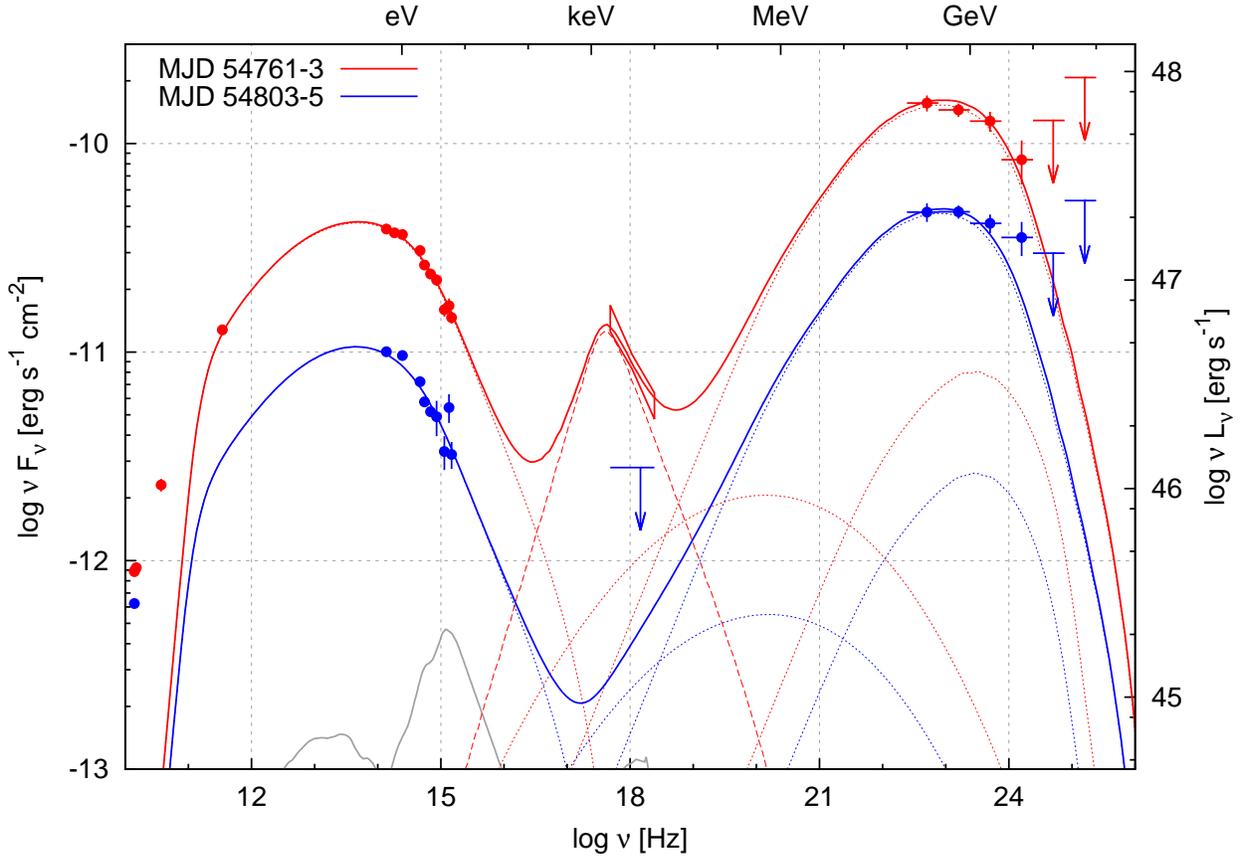}
  \caption{Numerical models fitted to observed spectral
states of AO~0235+164, dominated energetically by the
Comptonization of the infrared radiation from the dusty
torus (ERCIR). \emph{Red lines} show a fit to the high
state (MJD 54761-3), including the bulk-Compton feature
(\emph{dashed line}). \emph{Blue lines} show a fit to
the low state (MJD 54803-5). \emph{Dotted lines} indicate
individual spectral components, in order of increasing peak
frequencies: synchrotron, SSC, ERCIR, ERCBEL. \emph{Solid
lines} show the sums of all individual components. Note that presented  models do not cover the radio production which at $\nu < 100$GHz is strongly synchrotron-self-absorbed for our source parameters and must originate at much larger distances from the BH than a few parsecs. The \emph{gray line} shows the quasar composite SED adopted from \citet{key:elvis94} and normalized to the accretion disk luminosity $L_d = 4\times10^{45}$~erg~s$^{-1}$.}
  \label{fig:sed_overall}
\end{figure}

In this Section, we verify the ERC scenario by fitting
the observed SEDs with one-zone leptonic models \citep{key:mod03}.
We follow the evolution of relativistic electrons injected into a
thin spherical shell propagating conically with a constant
Lorentz factor $\Gamma_{\rm j}$ undergoing adiabatic and radiative
losses due to the synchrotron and inverse-Compton emission.
The external radiation includes
broad emission lines of characteristic photon energy
$E_{\rm BEL}\sim 10\;{\rm eV}$ and infrared dust radiation
of characteristic energy $E_{\rm IR}\sim 0.3\;{\rm eV}$.
We attempted to fit the high state of MJD 54761-3 with
a 'blazar zone' located either within (ERCBEL model) or
outside the BLR (ERCIR model).
In the ERCBEL model, the electron break inferred from
the synchrotron spectrum is too low to reproduce the $\gamma$-ray
spectrum above  $\sim 1$ GeV. This problem is absent in the
ERCIR model (red lines in Figure \ref{fig:sed_overall}).  This is because
Comptonization of IR photons is subject to much weaker
Klein-Nishina suppression in the GeV band than Comptonization of
optical/UV emission-line photons. The parameters
of the ERCIR model are: location $r=r_{\rm IR}$, Lorentz factor
$\Gamma_{\rm j}=20$, opening angle
$\theta_{\rm j}=1/\Gamma_{\rm j}=2.9^\circ$ (hence the Doppler factor $\mathcal{D}_{\rm j}=\Gamma_{\rm j}$), magnetic field strength
$B'=0.22\;{\rm G}$, viewing angle $\theta_{\rm obs}=2.3^\circ$.
Electrons are injected with a doubly-broken energy distribution
with $\gamma_{\rm br,1}=100$, $\gamma_{\rm br,2}=5800$, $p_1=1.5$, $p_2=2.03$, $p_3=3.9$.

The rate of electron energy injection is
$\dot{E}_{\rm e,inj}'\sim 4.8\times 10^{43}\;{\rm erg\,s^{-1}}$.
Over co-moving time
$\Delta t'\sim r_{\rm IR}/(2\Gamma_{\rm j}\beta_{\rm j}c)\sim 4\times 10^6\;{\rm s}$,
the total injected electron energy is
$E_{\rm e,inj}'\sim \dot{E}_{\rm e,inj}'\Delta t'\sim 1.9\times
10^{50}\;{\rm erg}$. At the end of the injection the total
number of electrons is $N_{\rm e}=6.6\times 10^{54}$ and their
total energy in the co-moving frame $E_{\rm e}'\sim 1.1\times
10^{50}\;{\rm erg}$. The average efficiency of electron energy
losses is $\eta_{\rm e,loss}=1-(E_{\rm e}'/E_{\rm e,inj}')\sim 0.42$.
The electron flux is $\dot{N}_{\rm e}\sim \pi\Gamma_{\rm j}R^2c
N_{\rm e}/V'\sim 1.2\times 10^{49}\;{\rm s^{-1}}$, where
$V'\sim 4\pi R^3/3$ is the volume of the emitting region in the co-moving frame
and $R\sim\theta_{\rm j}r$ is the jet radius. The electron energy
flux is $L_{\rm e}\sim\pi\Gamma_{\rm j}^2R^2c E_{\rm e}'/V'\sim 4.1\times
10^{45}\;{\rm erg\,s^{-1}}$ and the proton energy flux is
$L_{\rm p}\sim\pi\Gamma_{\rm j}^2R^2c N_{\rm p}m_pc^2/V'\sim 3.6\times
10^{47} \, (n_p/n_e) \;{\rm erg\,s^{-1}}$, where
$N_{\rm p}\sim N_{\rm e}(n_p/n_e)$ is the total number of protons
and $(n_e/n_p)$ is the lepton-to-proton number ratio.
The magnetic energy flux is $L_{\rm B}=\pi \Gamma_{\rm j}^2R^2c u_{\rm B}'\sim 4.4\times 10^{45}\;{\rm erg\,s^{-1}}$.
The resulting jet magnetization parameter is $\sigma_{\rm B}\sim L_{\rm B}/L_{\rm p}\sim 0.012 (n_e/n_p)$,
and the radiative efficiency is $\eta_{\rm rad,j}\sim L_{\rm obs}/(2\Gamma_{\rm j}^2L_{\rm p})\sim 0.022(n_e/n_p)$,
where $L_{\rm obs}\sim 6.4\times 10^{48}\;{\rm erg\,s^{-1}}$
is the observed bolometric luminosity of the source. The
relation between the jet power and accretion disk luminosity
is $L_{\rm p}/L_{\rm d}\sim 91(n_p/n_e)$. Parameterizing the jet
production efficiency by $L_{\rm p}\sim\eta_{\rm j}\dot M_{\rm acc}c^2$
and the radiative efficiency of the accretion disk
$L_{\rm d}\sim \eta_{\rm rad,d}\dot M_{\rm acc}c^2$, where
$\dot M_{\rm acc}$ is the accretion rate, we obtain
$n_e/n_p\sim 91(\eta_{\rm rad,d}/\eta_{\rm j})$. For $\eta_{\rm rad,d}\sim 0.1$
and $\eta_{\rm j}\sim 1$, we have $n_e/n_p\sim 9.1$ and hence
$\sigma_{\rm B}\sim 0.11$ and $\eta_{\rm rad,j}\sim 0.2$.
We are thus able to match the jet power with the accretion power onto the central black hole, adopting a reasonably low jet magnetization, which allows formation of strong shock waves.
The ratio of electrons to protons is consistent with the results of \cite{key:SM00}.
This model predicts the observed variability time scale
$ t_v \simeq (1+z)R/(c \Gamma_{\rm j}) \sim 8$ days,
consistent with the time scale of the significant variations of the optical and $\gamma$-ray flux.

As we noted above, the X-ray spectrum, at least during the first of the two epochs
considered here, is too soft to be interpreted as an SSC component and cannot be
the high-energy tail of the synchrotron component since it does not lie on the
extrapolation of the optical-UV spectrum (but see the caveats above, related to
the corrections for extinction).  Instead, it
can be explained by Comptonization of external radiation by a  population of
relatively cold electrons \citep{key:BS87,key:ravasio03}.  Such bulk Compton radiation
is expected to be produced in a jet much closer
to the black hole than the nonthermal blazar radiation, at distances at which
cooling of even mildly relativistic electrons is very efficient.
There the jet is still in the acceleration phase and, therefore, its bulk
Lorentz factor is expected to be smaller than in the blazar zone.
But noting that according to magnetohydrodynamical models the acceleration process is very
smooth (see, e.g., \citealt{key:komissarov}) and that the bulk Compton radiation must be significantly Doppler
boosted to be visible in the blazar spectra,  the dominant contribution
to bulk-Compton radiation is expected to be produced at distances
which are already well separated from the base of the jet
\citep{key:sikora05,key:celotti07}.
We consider a stationary emitting region at characteristic
radius $r_{\rm b}\sim 100 R_{\rm g}\sim 6\times 10^{15}\;{\rm cm}$,
where $R_{\rm g}=GM_{\rm BH}/c^2$ is the gravitational radius of the
central black hole of mass $M_{\rm BH}\sim 4\times 10^8\;M_\odot$.
The bulk Lorentz factor is $\Gamma_{\rm b}\sim 10$ and the Doppler factor is
$\mathcal{D}_{\rm b}=1/[\Gamma_{\rm b}(1-\beta_{\rm b}\cos\theta_{\rm obs})]\sim 16$.
Bulk-Compton luminosity is given by the approximate formula
\begin{equation}
L_{\rm b} \simeq  N_{\rm e,b}  |\dot E_{\rm e,IC}|_{\rm b}\frac{\mathcal{D}_{\rm b}^3}{\Gamma_b}\,,
\end{equation}
where $N_{\rm e,b}$ is the number of electrons enclosed in the
$\Delta r \sim r_{\rm b}$ portion of a jet,
$|\dot E_{\rm e,IC}|_{\rm b} = (4/3) c \sigma_T u_{\rm ext,b} \Gamma_{\rm b}^2$
is the rate of production of Compton radiation  by a single electron,
and $u_{\rm ext,b} = \xi_{\rm b} L_{\rm d} /(4\pi r_{\rm b}^2 c)$ is
the energy density of external radiation field,
which at distances $\le 100R_{\rm g}$ is very likely to be dominated
by rescattering of disk radiation by electrons in the accretion
disk corona of covering factor $\xi_{\rm b}$.
The energy spectrum of the external radiation is approximated
by a broken power-law distribution $u_{\rm ext,b}(E)\propto E^{-\alpha_{\rm i}}$ with $\alpha_1=0$, $\alpha_2=1.8$ and $E_{\rm br}=10\;{\rm eV}$ \citep{key:rich06,key:shang11}.
The electron flux is $\dot{N}_{\rm e,b}\sim N_{\rm e,b}c/r_{\rm b}$.
Assuming that it matches the electron flux in the blazar zone
($\dot{N}_{\rm e,b}\sim \dot{N}_{\rm e}$), we calculate the total
number of electrons producing the bulk-Compton component to be
$N_{\rm e,b}\sim 2.4\times 10^{54}$.  We find that the X-ray spectrum
of luminosity $L_{\rm b} \sim 8.7\times 10^{46}\;{\rm erg\,s^{-1}}$
can be reproduced with cold electrons for
$\xi_{\rm b} \simeq 0.19\,(r_{\rm b}/6 \times 10^{15}\,{\rm cm})$.

Multi-wavelength light curves show that X-rays do not correlate
with radiation in other spectral bands. This suggests  that X-ray
variability of the bulk-Compton radiation can be caused by local
 wiggling of the jet, e.g. caused by variations of the average direction of non-axisymmetric
outflows generated near the BH. Jet wiggling can also explain
independent variability in the blazar zone. We have fitted the
low state (MJD 54803-5) with an ERCIR model (blue lines in Figure \ref{fig:sed_overall})
very similar to the one for the high state, changing only
the viewing angle, from $\theta_{\rm obs}=2.3^\circ$ to
 $\theta_{\rm obs}=3.7^\circ$, and the magnetic field strength,
from $B'=0.22\;{\rm G}$ to $B'=0.20\;{\rm G}$.
With the new viewing angle, the observer
is placed outside the jet opening cone and the observed luminosity
decreases due to a lower effective Doppler factor. The change in
the magnetic field strength reduces the synchrotron luminosity by $\sim 30\%$
relative to the ERC luminosity. The overall
spectral shape is matched without any adjustment in the electron
energy distribution.

\section{Discussion}
\label{sec:discussion}

Multi-wavelength observations of blazars, including \objectname{AO~0235+164},
show that events associated with periods of greater activity occur
over the entire electromagnetic spectrum, from radio to $\gamma$-rays.
Light curves taken in different spectral bands correlate on time scales longer than a month,
while on shorter time scales correlations are weaker and some lags are claimed.
In particular, monthly lags of the radio signals following
the $\gamma$-rays are observed \citep{key:push10}.
This is interpreted in terms of the synchrotron opacity at radio frequencies
and indicates that $\gamma$-rays are produced at distances from the BH
that are several parsecs smaller than the position of the radio cores.
Whether $\gamma$-ray emission is smoothly distributed over several decades of distance \citep{key:BL95},
or is associated with specific locations of energy dissipation in a jet, is still debated.
The localized dissipation zones could be related to: reconnection
of magnetic fields which may operate efficiently at distances
$<0.01\;{\rm pc}$,
where the magnetization parameter $\sigma$ is expected to be large
\citep{key:nal11}; internal shocks resulting from collisions
between the jet portions moving with different velocities \citep{key:spada01},
which become
efficient at distances at which $\sigma$ drops below $0.1$;
oblique/reconfinement shocks which are formed at distances at which
interactions of a supersonic jets start to feel the influence of the external medium
\citep{key:hughes11,key:DM88,key:KF97,key:NS09}.

Using the SED of \objectname{AO~0235+164}, we showed in Section \ref{sec:modeling} that
the spectrum of this object can be reproduced using a model where
the dissipation takes place at a distance of $r \sim 1.7\;{\rm pc}$
from the BH and production
of $\gamma$-rays is dominated by Comptonization of NIR
radiation of hot dust. The activity of \objectname{AO~0235+164} during the same epoch
was analyzed independently by \cite{key:agudo11}. They included the VLBI
imaging data, and concluded that the
$\gamma$-ray emission zone is associated with the 7~mm radio core (i.e. at $\sim 12\;{\rm pc}$
from the BH) and propose that production of $\gamma$-rays is dominated by a turbulent multi-zone SSC process.
We comment on these differences below, noting that
we perform detailed modelling of the broad-band spectrum of the object.

As the long term radio light curves of \objectname{AO~0235+164} indicate, the active season in 2008 started
about 200 days before reaching the maximum. During this period
of time any portion of the jet propagates over a distance of
$\Delta r \sim c \Gamma^2 t_{obs}/(1+z) \ge$ tens of parsecs, i.e. much
larger than the length of a 'blazar zone' inferred from the time scales
of the short term flares. Such flares are very likely to be produced by
inhomogeneities of the
flow, radiating when passing through the region where energy dissipation
is intensified.
Particularly prominent flares are seen in the optical light
curves. Their $\sim 10$-day time scales imply that the extension of the dissipative
zone is
\begin{equation}
\Delta r \sim {c t_{fl} \over (1+z)(1 - \beta \cos{\theta_{obs}})}
\sim 1.7\;{\rm pc} \left(t_{fl} \over 10\;{\rm days}\right)
\left(\Gamma \over 20\right)^2 \left({\cal D}/\Gamma\right)\,,
\end{equation}
which nicely corresponds with the location of the blazar zone derived from our model.
In order to form such  flares at a distance of $12\;{\rm pc}$,
a Lorentz factor of $\Gamma \sim 50$ is required. Interestingly, a similarly
large Lorentz factor is required in the \cite{key:agudo11} model to explain
the observed $t<20$-day time scale of the flux decay at $\lambda=1\;{\rm mm}$.
This can be inferred by taking into account that radiative cooling of electrons
emitting at 1 mm is inefficient and that the  time scale
of the flux decrease, as determined by the adiabatic losses, is
$t_{ad} \simeq (R/c)(1+z) / (\theta_j \Gamma {\cal D})$.
VLBI observations of \objectname{AO~0235+164} do not exclude such a large value
of the bulk Lorentz factor \citep{key:jorstad,key:piner06}.

Finally we comment about the objections made by \cite{key:agudo11}
regarding the application of ERC models for the production
of $\gamma$ rays in \objectname{AO~0235+164}.
They pointed out that in this model it is impossible to explain
the lack of correlation between short-term variations of the
$\gamma$-ray and optical fluxes because of a lack of variations
of the external radiation field.
However, in the fast cooling regime, variations of the
inverse-Compton flux are determined not by variations
of the seed radiation field but by variations of
the electron injection function.  It does not matter
whether the seed radiation is external or internal,
so this criticism may also be applied to the SSC models.
Hence, the lack of a clear correspondence between variations
in these two spectral bands must have a different origin than
fluctuations in the background radiation.  They can be related
to variations of the Doppler factor and magnetic fields
in the  kinematically and geometrically complex dissipative zone.
In particular, this can be the case if such a zone is associated
with the oblique and/or reconfinement shocks, which in \objectname{AO~0235+164} is indicated
by roughly perpendicular orientation of the optical EVPA with respect to the jet
axis (see Section \ref{sec:polarization}).

\section{Conclusions}
\label{sec:conclusions}

\textit{Fermi}-LAT detected enhanced activity in the high-redshift BL Lac
object \objectname{AO~0235+164} during the first 6 months of operations.
We present the results of an intensive multi-wavelength campaign
covering radio, mm, near-IR, optical, UV and X-ray bands,
as well as optical polarimetry.
Extinction in the optical/UV/X-ray band, complicated by the existence of an additional absorbing system at intermediate redshift, has been carefully taken into account.
We proposed a modification to the extinction model introduced by \cite{key:junkkarinen} and used by \cite{key:raiteri2005} that corrects a spurious spectral feature in the FUV band.

The $\gamma$-ray spectrum is consistent with a broken power-law.
Hints of spectral variability can be seen in episodic increases
of the (1-100 GeV)/(0.1-1 GeV) hardness ratio. The brightest
$\gamma$-ray flare is much more pronounced in the 0.1-1 GeV energy band.

The $\gamma$-ray activity is roughly correlated with the activity
in the optical/near-IR band. There is a possible delay of 15 days
of the R-band flux with respect to the $\gamma$-ray flux.
The optical flux is also correlated with the optical polarization
degree, which reaches values up to 35\%. At the same time, the optical
polarization angle is close to $100^\circ$ with moderate scatter.
As is typical for blazars, the activity in the radio band
is smoother and begins months before the optical/$\gamma$-ray activity, while the radio-flux peaks are delayed by several weeks with respect to the higher energy bands.

The behavior of the source in the X-ray band is distinct from other bands,
as it shows a 20-day high state delayed by a month from the main optical/$\gamma$-ray flare.
The X-ray spectrum during the high state is unusually soft, $\Gamma\sim 2.6$,
and is inconsistent with the extrapolation of the optical/UV spectrum,
unless we assume a much stronger extinction.
We interpret this X-ray component as the bulk-Compton emission,
i.e. Comptonization of the accretion-disk radiation reprocessed
at the distance of $\sim 100\;R_{\rm g}$,
in the region of ongoing jet acceleration and collimation.
Such a feature has been tentatively reported before in a few sources,
however the present case is still not definitive.
The short duration of the high X-ray state can be explained by
a rapid ``wiggling'' of the inner jet.

The broad-band SEDs extracted for two different activity states are,
with the exception of the X-ray feature,
typical for luminous blazars.
We interpret the broad-band SEDs in the standard leptonic scenario,
with the low-energy bump due to synchrotron radiation
and the high-energy bump due to Comptonization of the
external infrared radiation from the dusty torus (ERCIR).
The energetic constraints are very tight,
because, if the jet power is comparable to the Eddington luminosity of the central black hole,
the required radiative efficiency of the jet is $\sim 20\%$,
the magnetization is $\sigma_{\rm B}\sim 11\%$
and the pair-to-proton ratio is $n_{\rm e}/n_{\rm p}\sim 9$.
The bulk Compton feature in the high X-ray state requires,
if the electron number flux is to be matched to the model of the flaring state,
a covering factor of the accretion disk corona $\xi_{\rm b}\sim 19\%$.
An alternative interpretation of the high-energy bump with the SSC emission requires a very low covering factor for the dusty torus, in conflict with the observations of quasars.

\begin{acknowledgments}

The \textit{Fermi} LAT Collaboration acknowledges generous ongoing support
from a number of agencies and institutes that have supported both the
development and the operation of the LAT as well as scientific data analysis.
These include the National Aeronautics and Space Administration and the
Department of Energy in the United States, the Commissariat \`a l'Energie Atomique
and the Centre National de la Recherche Scientifique / Institut National de Physique
Nucl\'eaire et de Physique des Particules in France, the Agenzia Spaziale Italiana
and the Istituto Nazionale di Fisica Nucleare in Italy, the Ministry of Education,
Culture, Sports, Science and Technology (MEXT), High Energy Accelerator Research
Organization (KEK) and Japan Aerospace Exploration Agency (JAXA) in Japan, and
the K.~A.~Wallenberg Foundation, the Swedish Research Council and the
Swedish National Space Board in Sweden.

Additional support for science analysis during the operations phase is gratefully acknowledged from the Istituto Nazionale di Astrofisica in Italy and the Centre National d'Etudes Spatiales in France.

We acknowledge the support by the Polish MNiSW grant N N203 301635.

L. C. Reyes acknowledges support from NASA through \textit{Swift} Guest Investigator Grant NNX10AJ70G; as well as support by the Kavli Institute for Cosmological Physics at the University of Chicago through grants NSF PHY-0114422 and NSF PHY-0551142 and an endowment from the Kavli Foundation and its founder Fred Kavli.

This research is partly based on observations with the 100-m telescope of the
MPIfR (Max-Planck-Institut f\"ur Radioastronomie) at Effelsberg.
This work has made use of observations with the IRAM 30-m telescope.

This paper is partly based on observations carried out at the German-Spanish Calar Alto Observatory, which is jointly operated by the MPIA and the IAA-CSIC.

The Abastumani team acknowledges financial support by the Georgian National Science Foundation through grant GNSF/ST08/4-404.

The Mets\"ahovi team acknowledges the support from the Academy of Finland
to our observing projects (numbers 212656, 210338, 121148, and others).

The Submillimiter Array is a joint project between the Smithsonian Astrophysical Observatory and the Academia Sinica Institute of Astronomy and Astrophysics and is funded by the Smithsonian Institution and the Academia Sinica.

The acquisition and analysis of the SMARTS data are supported by
Fermi GI grants 011283 and 31155 (PI C. Bailyn).

Data from the Steward Observatory spectropolarimetric monitoring project
were used. This program is supported by Fermi Guest Investigator grants
NNX08AW56G and NNX09AU10G.

UMRAO research is supported by a series of grants from the NSF and  NASA, most recently AST-0607523 and
Fermi GI grants NNX10AP16G and NNX11AO13G respectively; funds for  telescope operation are provided by
the University of Michigan.

\end{acknowledgments}

\end{document}